\documentclass[a4paper, amsfonts, amssymb, amsmath, reprint, showkeys,nofootinbib,twoside,unsortedaddress,superscriptaddress,floatfix]{revtex4-1}
\usepackage[english]{babel}
\usepackage[utf8]{inputenc}
\usepackage[colorinlistoftodos, color=green!40, prependcaption]{todonotes}
\usepackage{breqn}

\usepackage{amsthm}
\usepackage{mathtools}
\usepackage{physics}
\usepackage{xcolor}
\usepackage{graphicx}
\usepackage[left=23mm,right=13mm,top=35mm,columnsep=15pt]{geometry} 
\usepackage{adjustbox}
\usepackage{placeins}
\usepackage[T1]{fontenc}
\usepackage{lipsum}
\usepackage{csquotes}

\newcommand{\Lik}{\mathcal{L}}
\usepackage{booktabs}
\usepackage{array}

\begingroup\makeatletter
\catcode`,=\active
\global\let\breqn@comma,
\protected\gdef,{\ifmmode\expandafter\breqn@comma\else\expandafter\active@comma\fi}
\endgroup

\usepackage[pdftex, pdftitle={Article}, pdfauthor={Author}]{hyperref} %

\begin{document}
\title{Relevance of jet magnetic field structure for blazar axionlike particle searches}

\author{James Davies}
\email[Email address: ]{james.davies2@physics.ox.ac.uk}
\affiliation{University of Oxford, Department of Physics, Oxford OX1 3RH, United Kingdom}

\author{Manuel Meyer}
\email[Email address: ]{manuel.e.meyer@fau.de}
\affiliation{Erlangen Centre for Astroparticle Physics, University of Erlangen-Nuremberg, Erlangen 91058, Germany}

\author{Garret Cotter}
\email[Email address: ]{garret.cotter@physics.ox.ac.uk}
\affiliation{University of Oxford, Department of Physics OX1 3RH, Oxford, United Kingdom}

\date{\today} 

\begin{abstract}
Many theories beyond the Standard Model of particle physics predict the existence of axionlike particles (ALPs) that mix with photons in the presence of a magnetic field. One prominent indirect method of searching for ALPs is to look for irregularities in blazar gamma-ray spectra caused by ALP-photon mixing in astrophysical magnetic fields. This requires the modeling of magnetic fields between Earth and the blazar. So far, only very simple models for the magnetic field in the blazar jet have been used. Here we investigate the effects of more complicated jet magnetic field configurations on these spectral irregularities, by imposing a magnetic field structure model onto the jet model proposed by Potter $\&$ Cotter. We simulate gamma-ray spectra of Mrk 501 with ALPs and fit them to no-ALP spectra, scanning the ALP and B-field configuration parameter space and show that the jet can be an important mixing region, able to probe new ALP parameter space around $m_a\sim$ 1-1000 neV and $g_{a\gamma}\gtrsim$ $5\times10^{-12}$ $\text{GeV}^{-1}$. However, reasonable (i.e. consistent with observation) changes of the magnetic field structure can have a large effect on the mixing. For jets in highly magnetized clusters, mixing in the cluster can overpower mixing in the jet. This means that the current constraints using mixing in the Perseus cluster are still valid.
\end{abstract}


\maketitle

\section{Introduction}
\label{sec:background}

\label{sec:preintro}
\begin{figure}
        \centerline{\includegraphics[width = 0.5\textwidth]{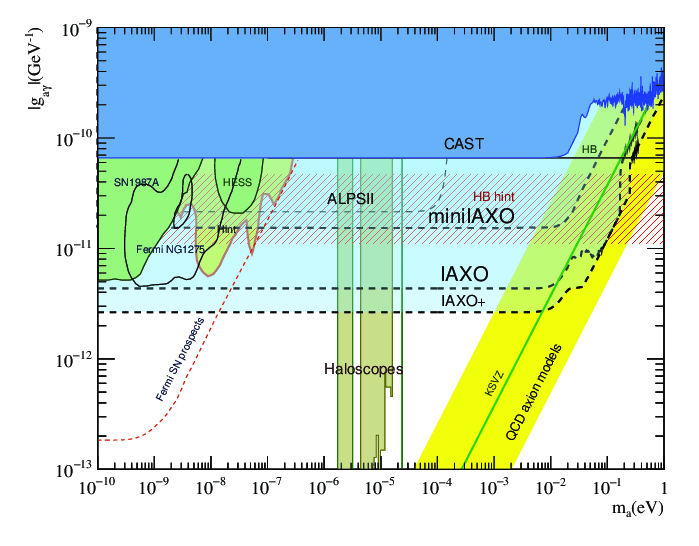}}
        \caption{ALP ($m_{a}$, $g_{a\gamma}$) parameter space with current and future predicted experimental exclusions. In particular, exclusions from CAST, IAXO, and Fermi observation of NGC 1275 and SMM SN1987A exclusions are shown (see text for details). The region where the axion could solve the CP problem is shown in yellow. Plot from Ref. \cite{exclusions}.}
        \label{exclusionplot}
    \end{figure}

The axion is a light ($m_{a}\lesssim 10$ meV \cite{particlereview,weinberg,wilczek}) pseudoscalar particle beyond the Standard Model which could theoretically solve the strong charge-parity (CP) problem \cite{peccei}. Importantly, axions couple to the electric field of photons ($\boldsymbol{E}$) in the presence of a magnetic field ($\boldsymbol{B}$). The Lagrangian
\begin{align}\label{eq:mixlagrangian}
    \Lik_{a\gamma}= -\frac{1}{4}g_{a\gamma}F_{\mu\nu}\tilde{F}^{\mu\nu}a = g_{a\gamma}\boldsymbol{E}\cdot\boldsymbol{B} a,
\end{align}
describes this coupling \cite{raffstod}, where $F^{\mu\nu}$ is the electromagnetic field tensor and $\tilde{F}^{\mu\nu}$ is its dual: $\tilde{F}^{\mu\nu}=\frac{1}{2}\epsilon^{\mu\nu\rho\sigma}F_{\rho\sigma}$, $g_{a\gamma}$ is the axion-photon coupling and $a$ is the axion field.
In order to solve the strong CP problem, the photon-axion coupling and the axion mass must be strongly related (\cite{peccei,particlereview}).
axionlike particles (ALPs) are generalizations of the axion where this coupling-mass relation is relaxed. While ALPs no longer generically solve the strong CP problem, they commonly arise in string theories and for certain masses and couplings, could, in theory, make up some or all of the dark matter content of the universe (e.g \cite{alpst,alpdm}).
The coupling between photons and ALPs leads to ALP-photon mixing in the presence of a magnetic field, meaning that photons can oscillate into ALPs and vice versa. This mixing has been the basis for many experimental searches for the axion (and ALPs) \cite{alpsearches} and, recently, a lot of indirect searches using astrophysical gamma-ray observations. So far, no ALPs have been found; Fig. \ref{exclusionplot} shows the experimental and observational exclusions across the ALP mass and coupling parameter space. 


Because of the strengths of astrophysical magnetic fields along the line of sight to gamma-ray sources, the region of ALP parameter space relevant for gamma-ray astronomy is $m_a \lesssim 10^{-6}$ eV. The most prominent experimental search here has been the CERN Axion Solar Telescope (CAST) experiment, which uses a large LHC prototype magnet to try to reconvert axions converted from photons in the core of the Sun back into photons, which can then be detected. The nonobservation of any ALPs by CAST has set an upper bound on the value of the coupling, $g_{a\gamma}$(95\% C.L.)$< 6.6 \times 10^{-11}\text{GeV}^{-1}$, for ALP masses $m_{a}<0.02$ eV \cite{cast,cast_new,alpsearches}. The International Axion Observatory (IAXO), a next generation helioscope, should be able to obtain a signal-to-noise ratio about 5 orders of magnitude better than CAST. IAXO should then be able to probe couplings $g_{a\gamma}$ as low as $5\times 10^{-12}\text{GeV}^{-1}$ for $m_{a}<0.01$ eV \cite{iaxo}.\par
As well as CAST, gamma-ray observations provide competitive limits on ALP mass and coupling in this region.
 Photons would be expected to convert to ALPs in core-collapse supernovae. This would mean a supernova would emit ALPs, some of which would be expected to reconvert into photons in the magnetic field of the Milky Way. The nonobservation of a gamma-ray signal simultaneous with the detected neutrinos from supernova SN1987A by the Solar Maximum Mission (SMM) satellite has constrained $g_{a\gamma}\lesssim 5.3\times10^{-12}\text{GeV}^{-1}$ for masses $m_{a}< 4.4 \times 10^{-10}$ eV \cite{supernova}. Recent limits from the absence of gamma-ray flashes detected by Fermi-LAT coincident with extragalactic supernovae are also similar \cite{manuel_supernova}.\par
ALP-photon mixing also suggests the indirect detection of ALPs by the spectral irregularities they would cause in smooth astrophysical gamma-ray spectra where magnetic fields are found along the line of sight to the source. For example, analysis of data from the Fermi Large Area Telescope (Fermi-LAT), looking for irregularities in the spectrum of NGC 1275, has excluded (95\% C.L.) values of the coupling $g_{a\gamma}>5\times 10^{-12}\text{GeV}^{-1}$ (as low as the planned sensitivity of IAXO) for masses $0.5 \lesssim m_{a} \lesssim 5$ neV (see Fig. \ref{exclusionplot}) \cite{manuelfermi}. This kind of search has also been done with other sources in the Fermi data, as well as with H.E.S.S. and x-ray telescopes (e.g. \cite{fermi_pks2155304,fermi_pulsars,hess_exclusions,chandra_ngc1275}). farther searches are also planned, with instruments such as the Cherenkov Telescope Array (CTA) \cite{science_with_cta,cta_performance}, which is soon to come online with a sensitivity that will be better than other gamma ray telescopes by an order of magnitude (e.g. \cite{cta_gpropa,galanti_blazars}).\par 
These searches require models of the magnetic fields along the line of sight. So far, all of these works have only used relatively simple blazar jet magnetic field models, either of a completely random domainlike structure \cite{sanchez-conde,harris,hochmuth_sigl} or of a completely ordered transverse field \cite{mena_razzaque,ronc_agn,galanti_blazars}. These simple models have been justified by the argument that most of the mixing occurs in other magnetic field environments along the way, such as the IGMF \cite{galanti_blazars} or a cluster magnetic field \cite{manuelfermi}, due to the small comparative size of the jet.\par
The focus of this work is to determine whether, and to what extent, the magnetic field structure of the blazar jet affects these oscillatory spectral features -- and hence, to determine the importance of blazar jet models for future ALP searches. After a brief overview of jets, in Sections \ref{subsec:jets} and \ref{sec:otherenvs} we discuss the modeling of blazar jets and other relevant magnetic field environments along the line of sight. To determine whether mixing in the jet is important compared to mixing elsewhere [IGMF, Milky Way, intracluster medium (ICM)], we simulate spectra of the blazar Mrk 501, including and excluding mixing in the jet (Section \ref{sec:matter}). Then, to determine how much the detailed structure of the jet matters, we define a jet model and scan the jet parameter space within the limits of current observation, simulation, and theory with many simulations of Mrk 501 spectra (Section \ref{sec:structure}). Then, we extend our results beyond Mrk 501 to blazars in general (Section \ref{subsec:beyond}).

\section{Modeling the Blazar Jet}
\label{subsec:jets}
Jets are ubiquitous across the gamma-ray sky, with active galactic nuclei (AGN) making up almost all the extragalactic gamma-ray sources \cite{4fgl}. Most regular galaxies harbour a super massive black hole (SMBH) at their core (\cite{eht_1,lynden-bell}). Some of these cores are active, emitting a lot of radiation as the SMBH accretes surrounding matter. About 1/10 of these AGN host strong jets: relativistic outflows of matter which propagate out to $\sim$Mpc scales, powered by the accretion flow and the black hole spin \cite{beg_bland_rees}. These jetted AGN make up the traditional radio galaxies, broadly categorized into two morphological classes: powerful Fanaroff-Riley type-II sources (FR IIs, edge brightened) having strong collimated jets inflating large lobes in the surrounding medium, and weaker FR I (core brightened) sources with unstable, plumelike jets unable to inflate lobes \cite{fr}. About 1/10 of these jetted AGN happen to be pointing their jets right at us. These are blazars, whose observed emission is strongly enhanced by relativistic boosting effects. Blazars are broadly split into flat spectrum radio quasars (FSRQs) and BL Lacs by their emission lines, roughly corresponding to FR IIs and FR Is respectively \cite{blandford_rev}.\par
These jets must be extremely powerful ($\sim 10^{46} \text{erg} \text{s}^{-1}$ \cite{beg_bland_rees}), producing stable relativistic flows that remain collimated out to very large distances. The details of the launching and collimation mechanism, particle content, particle acceleration mechanisms, and emission processes are all open questions -- though a reasonably coherent picture is starting to emerge. One prominent model for launching relativistic jets is the Blandford-Znajek mechanism, wherein magnetic field lines brought in by the accretion flow are wound up by the spinning black hole, extracting rotational energy from the black hole and creating magnetically dominated jets \cite{bz}. This magnetic energy can then be used to accelerate particles within the jet (by, e.g., reconnection, or magnetoluminescence) until rough equipartition between magnetic and particle energy is reached.\par
\begin{figure}
        \centerline{\includegraphics[width = 0.5\textwidth]{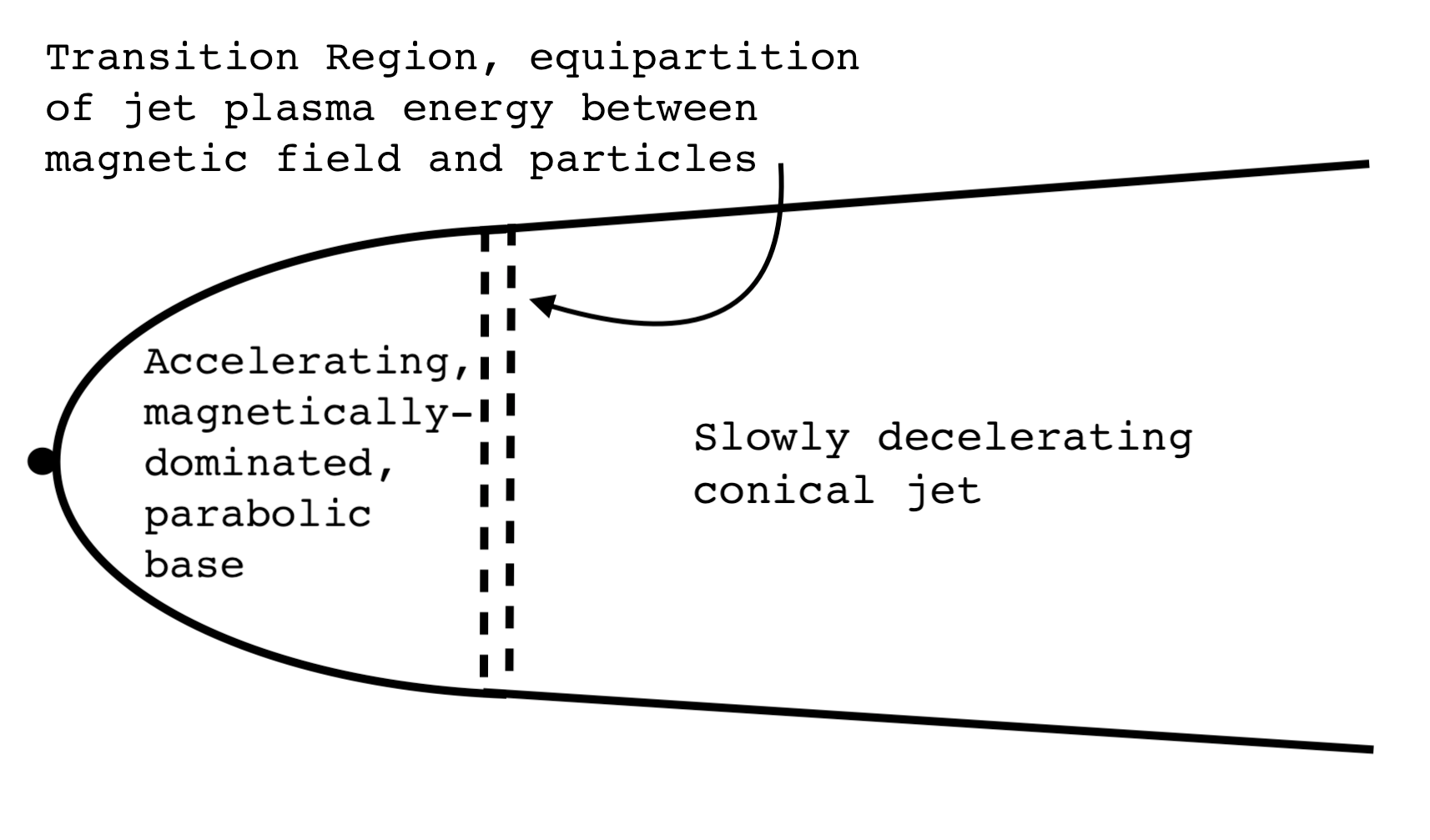}}
        \caption{Diagram showing the PC jet model: a magnetically dominated parabolic base that transitions to a conical ballistic jet once the magnetic field and particles come into energy equipartition. Gamma-ray emission is strongly dominated by the transition region (see Fig. \ref{fig:b_vfv}.)}
        \label{fig:pc_cartoon}
    \end{figure}

This idea, with an electron-positron jet, is used by Potter and Cotter (PC) to build a jet model, which is able to produce the full broad-band spectrum of many blazars, fitting observations remarkably well (see Refs. \cite{pc1,pc2,pc3,pc4,pc_nc}).The PC model is a 1D time-independent relativistic fluid flow. The structure of the PC jet model is an accelerating, parabolic, magnetically dominated base that transitions to a conical, slowly decelerating ballistic jet at around $10^5 \text{r}_g$ from the black hole, where $\text{r}_g$ is the gravitational radius of the black hole, which depends on its mass as $\text{r}_g=2GM/c^2$ (see Fig. \ref{fig:pc_cartoon}). The jet is populated by relativistic electrons and positrons (hereafter electrons), which come into rough energetic equipartition with the magnetic field at this transition region. Relativistic energy momentum and electron number are conserved down the jet, and the electrons are evolved with radiative and adiabatic losses. This electron population emits synchrotron radiation toward the radio end of the spectrum and Compton scatters its own synchrotron photon field, as well as others along the way to produce gamma rays. Depending on the source, photons from the accretion disk, broad line region, dusty torus, narrow line region, CMB, and starlight can all play a role in producing the gamma-ray emission. Emission is calculated all along the jet (it is not a one-zone model) and is found to fit the broadband spectra of many blazars.\par
A typical blazar spectrum is double peaked, with emission over the whole electromagnetic range, from radio to gamma rays \cite{emission_review}.
The current understanding is that the lower peak is produced by synchrotron radiation of electrons in the jet. The other peak, at high energy, is generally thought to be inverse Compton (IC) emission from the electrons either upscattering the synchrotron photons [synchrotron self-Compton, SSC] or other photon fields [external Compton, EC]. Because IC increases photon energy by a factor of $\gamma^2$, most of this gamma-ray emission would be expected to originate down near the base of the jet where bulk Lorentz factors are highest. Time-variability arguments also imply a fairly small gamma-ray emission region $\sim 10^5 - 10^6 r_g$ from the black hole \cite{blandford_rev}. Of course there will be some IC emission throughout the jet. As mentioned above, the PC model provides emission information for the whole jet, and they find that the transition region near the base of the jet strongly dominates the gamma-ray emission (see Fig. \ref{fig:b_vfv}). These simple leptonic (electron and positron) models of jets fit observations very well (see Ref. \cite{pc_nc}).\par
There are also hadronic models of jets, where relativistic protons play a role in producing the high energy peak from, e.g., proton-synchrotron, proton-pion production, and photo-pion production (e.g., \cite{petropoulou_psynch,muecke_had}). These models can also be made to fit observations but may suffer from energy considerations, often requiring super-Eddington power (it is much harder to accelerate a proton than an electron) \cite{bottcher_paradigm}. These models have recently received renewed attention due to the IceCube detection of very high energy neutrinos coincident with a gamma-ray flare from blazar TXS0506+056 \cite{icecube}, it being much easier to create neutrinos from protons than electrons. There no doubt must be some protons within the jet, but the question of how many and whether they play an important dynamical and radiative role is still up for debate (e.g., review \cite{blandford_rev}). Overall, the detailed processes of blazar gamma-ray emission are not well understood, and this is one reason why investigating the consequences of a PC-type model is important. This work is conducted in the context of the PC leptonic jet models.

\subsection{Jet magnetic field}
\label{subsec:jetfield}
The overall jet magnetic field strength and distance dependence are relatively well agreed upon. Minimizing the total energy in a synchrotron plasma that is emitting at a given synchrotron luminosity gives close to equipartition between particle and magnetic energy. Doing this for radio observations of jets gives strengths up to $\sim$ G near the base. VLBI Faraday rotation measure (RM) observations display a $B\propto1/r$ dependence down the jet \cite{osullivan_gabuzda}. VLBI observations of the very base ($\lesssim$ pc scales) of the very nearby jet of M87 find parabolic structure \cite{m87_vlbi}. GRMHD simulations also show a parabolic, magnetically dominated, accelerating jet base (e.g., \cite{mckinney_2006}). This parabolic structure is included in the PC model, which gives magnetic field strength all down the jet \cite{pc2}.
Because of the $\boldsymbol{E}\cdot\boldsymbol{B}$ term in $\Lik_{a\gamma}$, ALPs only see the component of external magnetic field \textit{transverse} to their direction, so we need to model the magnetic field direction as well as the strength along the whole of the jet to solve the mixing equations (Eq. \ref{eq:pgg} in App \ref{app:mixing}).
In contrast to the overall strength, the jet magnetic field orientation structure is currently not very well understood, especially at small scales close to the black hole. Conserving magnetic flux down a flux-frozen conical jet, one would expect the transverse component of magnetic field to be $\propto1/r$ and the component down the jet to be $\propto1/r^2$ \cite{beg_bland_rees}. Therefore, na\"ively, the overall magnetic field should become dominated by the transverse component with strength $\propto1/r$ (this is what is used in current ALPs work, e.g., \cite{galanti_blazars}). But, jet magnetic field structure is certainly more complicated, particularly at small radii.
For one thing, in a parabolic base, the same flux argument would make the transverse component go as $1/r^a$, where $a$ is the parabolic index ($\sim0.58$ in M87). VLBI RMs can also indicate the internal structure of the jet magnetic field as the RM depends on the integral of the field along the line of sight. Many rotation measure maps show asymmetry across the jet in a way indicative of helical magnetic fields at both parsec and (for powerful sources) kiloparsec scales \cite{gabuzda_helical,launchterm,larionov3c279} -- meaning the field is not purely transversal. A helical magnetic field would not be unusual near the base considering jets launched by something like the Blandford-Znajek mechanism where a poloidal field threading the black hole is wound up. In this scenario, a helical magnetic field might be expected with a pitch angle starting roughly as the ratio of jet flow and black hole rotation speeds and increasing down the jet as the jet expands and the magnetic energy is used to accelerate particles. Many simulations of relativistic jets show this helical field behaviour as well (e.g. Ref. \cite{mckinney_2009}).
It also seems unlikely that the magnetic field within such a violent environment would be entirely ordered. A tangled field could change the transverse field dominance, even at large radii, because a large poloidal field with many reversals could nevertheless have a small net flux down the jet. Entrainment of matter and the surrounding field would likely have a tendency to disorder the magnetic field, producing a tangled component (as well as changing the dependence from flux conservation). Fitting of models to Mrk 501 RM maps has provided observational support for this idea, deriving fractions of magnetic energy density in a tangled component up to $0.7$ \cite{gabuzda_mrk501}. So, for ALP searches, it is not obvious how to model the transverse magnetic field component in the jet properly.

 \section{Other Relevant Magnetic Field Environments}
 \label{sec:otherenvs}
 There are other magnetic field environments between us and blazars. In order to model the effects of ALP-photon mixing on astrophysical observations, magnetic field environments along the whole line of sight must be modeled. As well as the jet, we model the intergalactic magnetic field (IGMF) and the galactic magnetic field (GMF) of the Milky Way. Other searches have used a cluster magnetic field (CMF) as the main mixing region (e.g., \cite{manuelfermi}). For simplicity and generality, we do not at first model a cluster magnetic field as we are trying to find the general relevance of jet magnetic fields, and the cluster environment is very source dependent. We do, however, discuss the effects of a cluster field in Section \ref{subsec:beyond}.\par
 In modeling the IGMF, we follow much previous work (Refs. \cite{manuelfermi,galanti_blazars,ronc_smoothed} for instance) in having a randomly oriented, domainlike structure. Coherence lengths vary between $0.2$ Mpc and $10$ Mpc, with field strength of $\sim$ nG. This is about the strongest IGMF consistent with observations, and so will provide the strongest possible IGMF ALP-photon mixing \cite{igmf}. We want this as we are trying to tell whether there are regions of ALP parameter space for which mixing in the jet is important -- i.e. not just dominated by the IGMF -- so we need to model the strongest IGMF possible. We take the electron number density in the IGMF to be $n_e^{igmf}=1\times10^{-7}\text{cm}^{-3}$ as in Ref. \cite{galanti_blazars}.
 Another factor concerned with intergalactic space that affects the gamma-ray spectra is absorption from the extragalactic background light (EBL). The EBL is made up of starlight and light absorbed and reemitted by dust, integrated over the history of the universe. High-energy gamma rays can have enough energy to pair produce ($\gamma\gamma\rightarrow e^+e^-$) with photons of the EBL. This causes an exponential attenuation of the gamma-ray signal at high energies, essentially an optical depth \cite{dwekEBL}. We include EBL absorption as an energy-dependent mean free path in the photon propagation term of the mixing equation (Eq. \ref{eq:dgg} in App. \ref{app:mixing}). This absorption must be included as part of the ALP-photon beam equations of motion and cannot just be superimposed onto the spectrum afterward as ALPs do not suffer from EBL absorption, and so ALP-photon mixing changes the simple exponential form of the absorption. Interestingly, by traversing large portions of intergalactic space in an ALP state, high energy gamma rays could be received from much farther than their mean free path for pair production would seem to allow -- effectively reducing the opacity of the universe for gamma rays (see \cite{manuel_opacity} for a discussion of this). We use the EBL model of Dominguez et al. \cite{dominguezEBL}.\par
 For the GMF, we use the model of Janson and Farrar (\cite{jansonGMF}), which includes a disk field and a halo field with strength $\mathcal{O} \mu$G. We take the electron number density in the Milky Way to be $n_e^{mw}=1.1\times10^{-2}\text{cm}^{-3}$ as in Ref. \cite{galanti_blazars}.

\section{Relevance of Jet Magnetic Field Structure for ALP searches}
\label{thisyear}

\begin{figure*}
  \centering
    \includegraphics[width=1.0\textwidth]{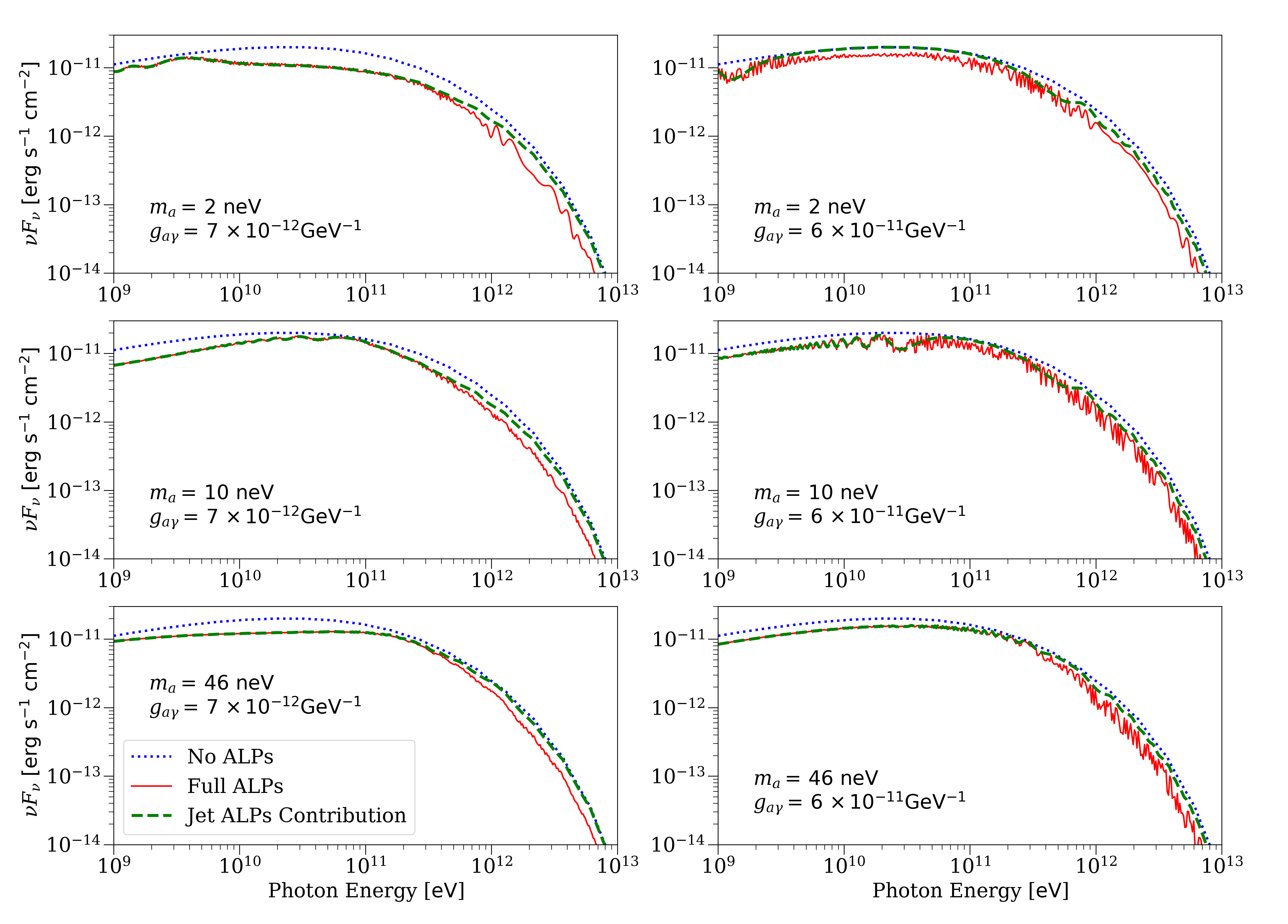}
  \caption{High energy peaks of Mrk 501 spectra, showing ALP-photon mixing induced oscillations for different values of ALP mass ($m_a$) and coupling ($g_{a\gamma}$). Blue (dotted) lines are the fit to data from the PC model. Red (solid) lines show the spectrum when mixing in the jet, IGMF, and Milky Way is included. Green (dashed) lines show the spectrum when only mixing in the jet is included.}
  \label{fig:spectra}
\end{figure*}

\subsection{The importance of mixing in the jet}
\label{sec:matter}
To investigate the relevance of more detailed jet structure for blazar ALP searches, it is first important to investigate whether the jet itself is important at all. As discussed above, So far, only very simple jet magnetic field models have been used for ALP searches because the IGMF or CMF have been assumed to dominate the mixing. For the jet to be important for ALP searches in general, there would have to be mixing in the jet that is not afterward washed out by the IGMF. In order to investigate this possibility, we simulate the effects of ALP-photon mixing on the gamma-ray spectrum of Mrk 501 (see App. \ref{app:mixing} for mixing equations). Mrk 501 is a very bright, fairly nearby ($z=0.034$) blazar whose magnetic field has been fairly well studied. It is also a candidate for upcoming CTA ALP searches and is already the subject of simulations in this area \cite{galanti_blazars}. \par
We use the standard ALP-photon mixing equations (see e.g. Refs. \cite{ronc_big} and \cite{ronc_smoothed} for a detailed discussion) an outline of which is given in Appendix \ref{app:mixing}. The oscillation length of ALP-photon mixing depends on the ALP mass ($m_a$), the ALP-photon coupling ($g_{a\gamma}$), the photon energy ($E$), the transverse field strength ($B_T$), and electron number density ($n_e$) of the mixing environment. The mixing equations can be used to propagate ALP-photon beams of various energies through a series of magnetic field environments and find the photon survival probability ($P_{\gamma\gamma}(E)$) at the end. The observed gamma ray spectrum from the source is then just the intrinsic spectrum multiplied by $P_{\gamma\gamma}(E)$. We follow the same procedure as previous work, with a slight alteration because a relativistic jet is not a cold plasma. Therefore, the photon effective mass depends on the nonthermal distribution function of electrons in the jet instead of simply being the plasma frequency as it is in a cold plasma (see Appendix \ref{app:mixing} for details).\par
Figure \ref{fig:spectra} shows some simulated spectral energy distributions (SEDs) of Mrk 501 for various ALP masses and couplings -- displaying the kind of spectral oscillations ALP-photon mixing can cause. The energy range we look at is 1 GeV - 10 TeV, roughly the important energy range for Mrk 501 with \textit{Fermi}-LAT and CTA \cite{cta_performance}. At this stage, we use the simple fully transverse jet magnetic field model of Refs. \cite{ronc_agn} and \cite{galanti_blazars}, as we are just trying to find whether the jet is relevant. That is, $B_T = B \propto 1/r$ with $B(r_{em}=0.3 \text{pc}) = 0.8$G and $n_e(r_{em}) = 5\times10^4\text{cm}^{-3}$, where $r_{em}$ is the location of the gamma-ray emission region measured from the black hole. The blue (dotted) lines are the gamma-ray peak of a a fit to broadband data\footnote{The data for this fit was taken from many telescopes at many frequencies (e.g. \textit{Fermi} for gamma rays, \textit{Swift} (UVOT, XRT, BAT) for x-ray to optical, RATAN-600 for radio, as well as others) and was collected in Ref. \cite{abdo_data} to obtain quasisimultaneous broadband spectra for sources in the \textit{Fermi}-LAT Bright AGN Sample.}, from the PC model taken from Ref. \cite{pc_nc}, without any ALP-photon mixing (hereafter ALP-less spectra).

\begin{figure}
  \centering
    \includegraphics[width=0.5\textwidth]{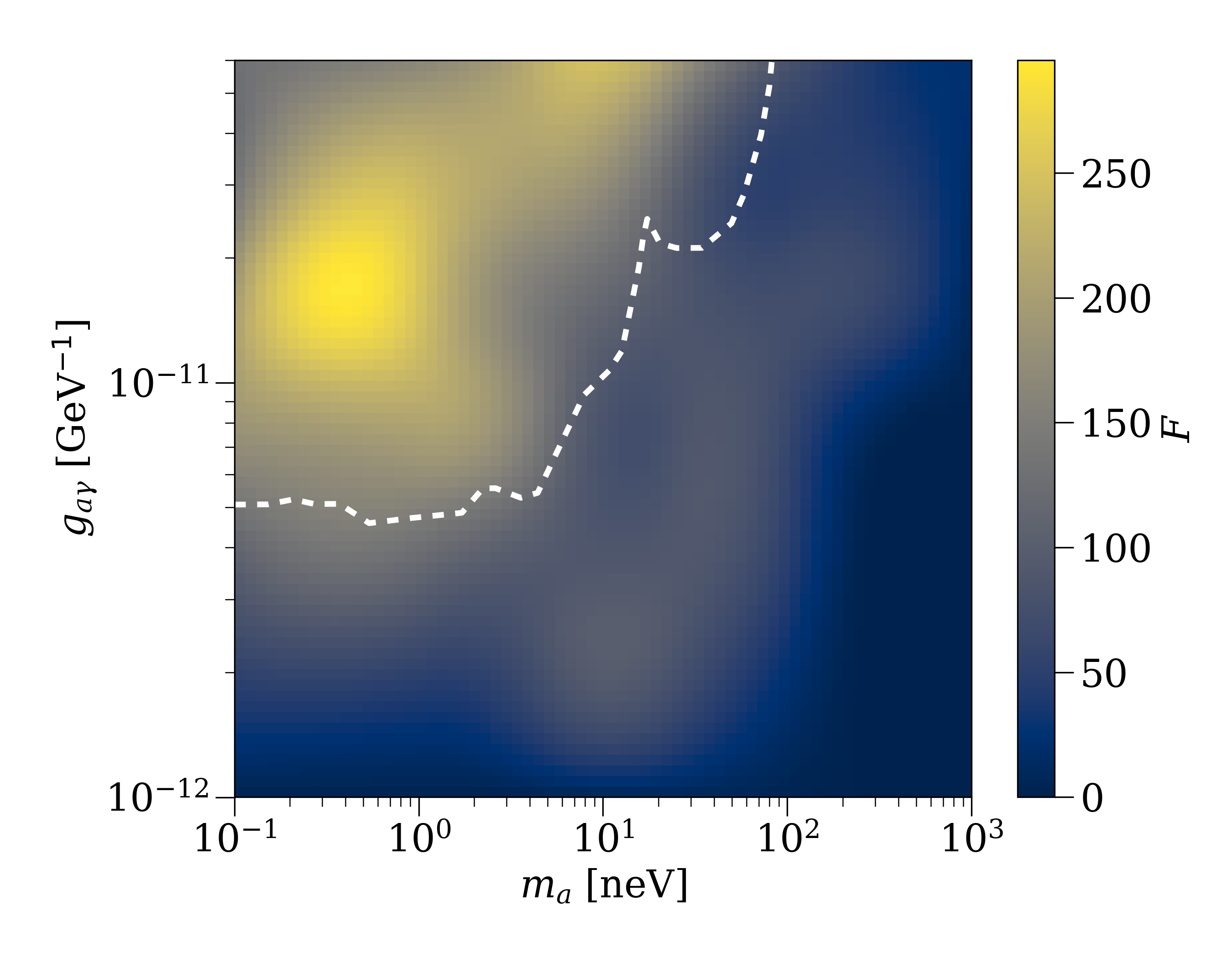}
  \caption{Values of $F$ -- sum of least-squares fit between ALP and ALP-less spectrum -- calculated over ALP mass-coupling parameter space. Dashed line shows current exclusions (see Sec. \ref{sec:preintro}).}
  \label{fig:fs}
\end{figure}

The green (dashed) lines show the SED produced by mixing only in the jet, and the red (solid) lines show the total effects of mixing in the jet, IGMF and Milky Way.
For ALP searches, the important property of these spectra are how well they can be distinguished from the ALP-less spectrum. In other words, for an ALP of a certain mass and coupling to be excluded (or discovered), it must produce large enough changes in the spectrum to be noticed. We use a least-square fit to quantify how distinguishable a given ALP spectrum is from the ALP-less spectrum (see Fig. \ref{fig:fs}). The value used is the minimum sum of the squared difference (least squares) between the ALP spectrum ($\phi^{alp}$) and ALP-less spectrum ($\phi$) across the relevant energy range, allowing the ALP spectrum to shift vertically\footnote{This is because a uniformly down-shifted ALP spectrum (i.e. uniformly strong mixing) with a higher intrinsic normalization is observationally indistinguishable from an ALP-less spectrum with lower normalization. In other words, we only care about changes in the \textit{shape} of the spectra.}: $F = \Sigma_i (\phi_i - \phi_i^{model})^2/0.01$. $F$ cannot be used to actually exclude ALP parameters because it is a comparison of idealised observed spectra with and without ALPs as opposed to a comparison of actual observed spectra. An actual ALP search would require folding the simulated intrinsic spectra with the IRF of a specific telescope and then using a log-likelihood method to compare expected counts with observed counts across the significant energy bins. Reference \cite{manuelfermi}, for example, uses this methodology with Fermi-LAT data to exclude a region of ALP parameter space. But, $F$ can be used to find what areas of parameter space could potentially be excluded. Also, by seeing how mixing in different regions contributes to the fits, we can tell how relevant each of the regions are for future searches in a general and instrument-independent way.

\begin{figure}
  \centering
    \includegraphics[width=0.5\textwidth]{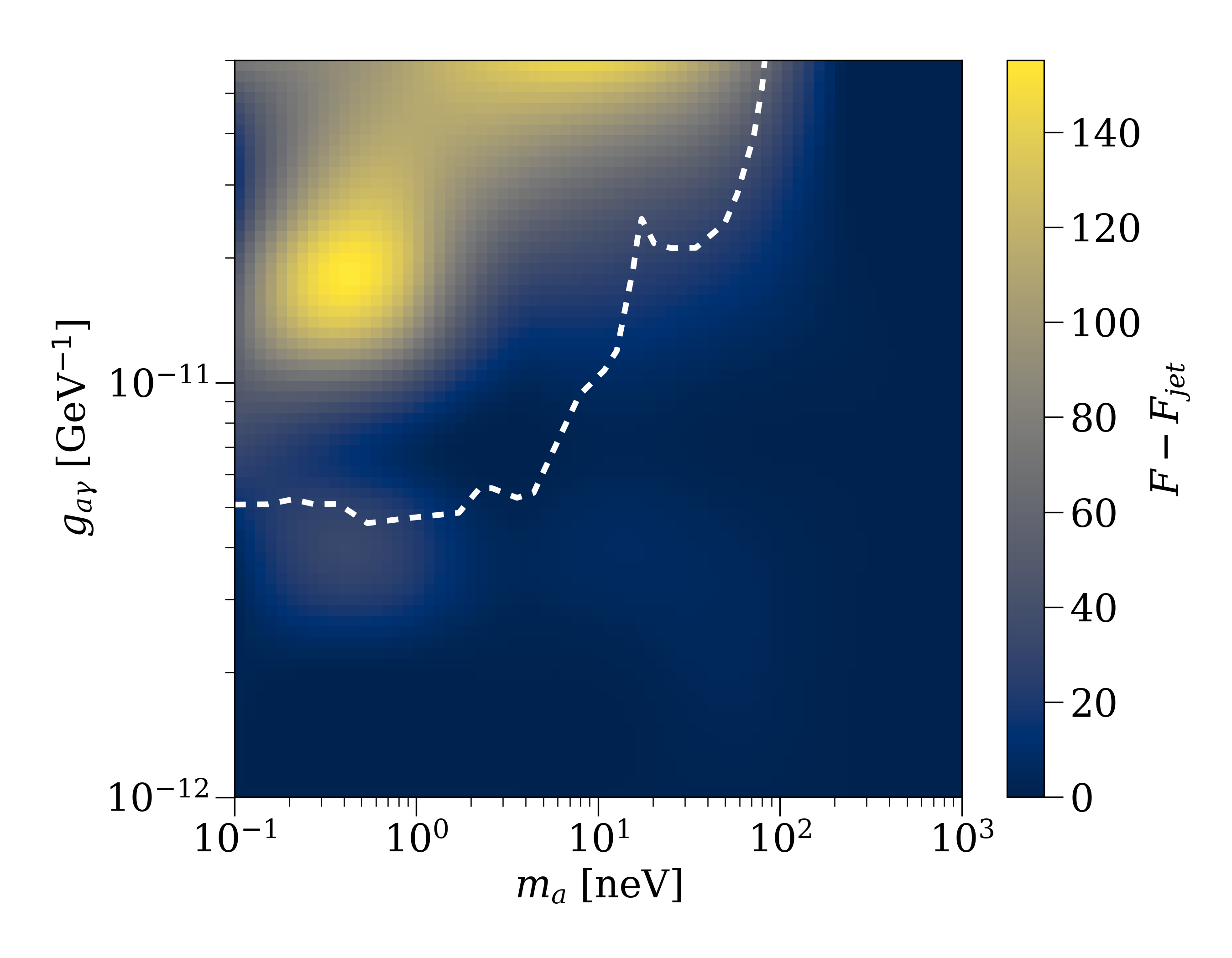}
  \caption{$F - F_{jet}$ over ALP mass-coupling parameter space. $F$ is the fit including mixing in the jet, IGMF, and the Milky Way. $F_{jet}$ is the fit only including mixing in the jet. Dashed line shows current exclusions (see Sec. \ref{sec:preintro}).}
  \label{fig:diffs}
\end{figure}

Figure \ref{fig:fs} shows the values of $F$ for the full (jet, IGMF, Milky Way) ALP spectra over the relevant ALP $(m_a,g_{a\gamma})$ parameter space. The white dashed contour shows the current exclusions (see Sec. \ref{sec:preintro}). The higher the value of $F$, the worse the fit between the ALP and ALP-less spectra, and so, the easier those parameters are to exclude; it makes sense that the highest values of $F$ are in the region of $(m_a,g_{a\gamma})$ space that has already been excluded. In principle, any pair of mass and coupling that produces a nonzero fit is accessible to future ALP searches of Mrk 501 and (as discussed in Sec. \ref{subsec:varying}) probably to BL Lac objects in general. It can be seen from Fig. \ref{fig:fs} that there is a fairly large region of unprobed parameter space with $F\sim100$, which is, therefore, accessible to future searches. Figure \ref{fig:diffs} shows $F - F_{jet}$ for the same region of $(m_a,g_{a\gamma})$ parameter space. $F_{jet}$ is the value of the $F$ when only mixing in the jet is included (i.e. without IGMF or MW mixing). A low value of $F-F_{jet}$ means that for this pair of mass and coupling mixing in the jet contributes most to the fits, and a high value means that the IGMF or the Milky Way dominates. For much of the parameter space, the jet contributes strongly to the mixing ($F-F_{jet}$ is small). In particular, the jet dominates the fits in the region outside the current exclusions that was shown in Fig. \ref{fig:fs} to be relevant for future searches. In other words, in order to extend ALP exclusions using blazar spectra, mixing in the jet has to be taken into account. The reason mixing in the jet has such a large effect on the fits (and, therefore, the possible exclusions) is that it sets the large scale shape of the spectrum, whereas mixing elsewhere causes small scale oscillations around this shape (compare green (dashed) and red (solid) lines in Fig. \ref{fig:spectra}). When fitting across the whole spectrum, small oscillations have a tendency to cancel out, whereas differences in large-scale shape do not. The difference in the oscillations caused by the two environments is because in the strong magnetic field of the jet, the propagation equations often enter the energy-independent strong mixing regime for large portions of the spectrum, with large oscillations occurring at the boundaries of this regime (see Appendix \ref{app:mixing}). In contrast, the IGMF is made up of many randomly oriented cells, meaning the transverse magnetic field can take on many values, leading to oscillations across the spectrum. For the region of ALP parameter space where the jet dominates, the difference in scale of the oscillations due to the IGMF and the jet can be intuitively understood by looking at the difference in length of the two regions compared to the average ALP-photon oscillation length within them. Intergalactic space is very large compared to the average oscillation length within it, so propagation in the IGMF goes over many oscillation lengths, leading to rapid oscillations in energy space. The jet is comparatively much smaller, so (even with the shorter average oscillation length within it) propagation through the jet goes over much fewer oscillation lengths, inducing comparatively slow oscillations in energy space. For example, for a pair of ALP parameters where the jet is important ($m_a = 40$ neV and $g_{a\gamma} = 6\times10^{-12}$ GeV$^{-1}$) and a photon energy of $E \sim 10$ GeV, the ALP-photon beam traverses around $10^5$ oscillation lengths in intergalactic space, but only about 35 within the jet. For lower masses, where the jet is not dominant, the situation can be different because the ALP-photon oscillation length depends on the ALP parameters.

\begin{figure}
  \centering
    \includegraphics[width=0.5\textwidth]{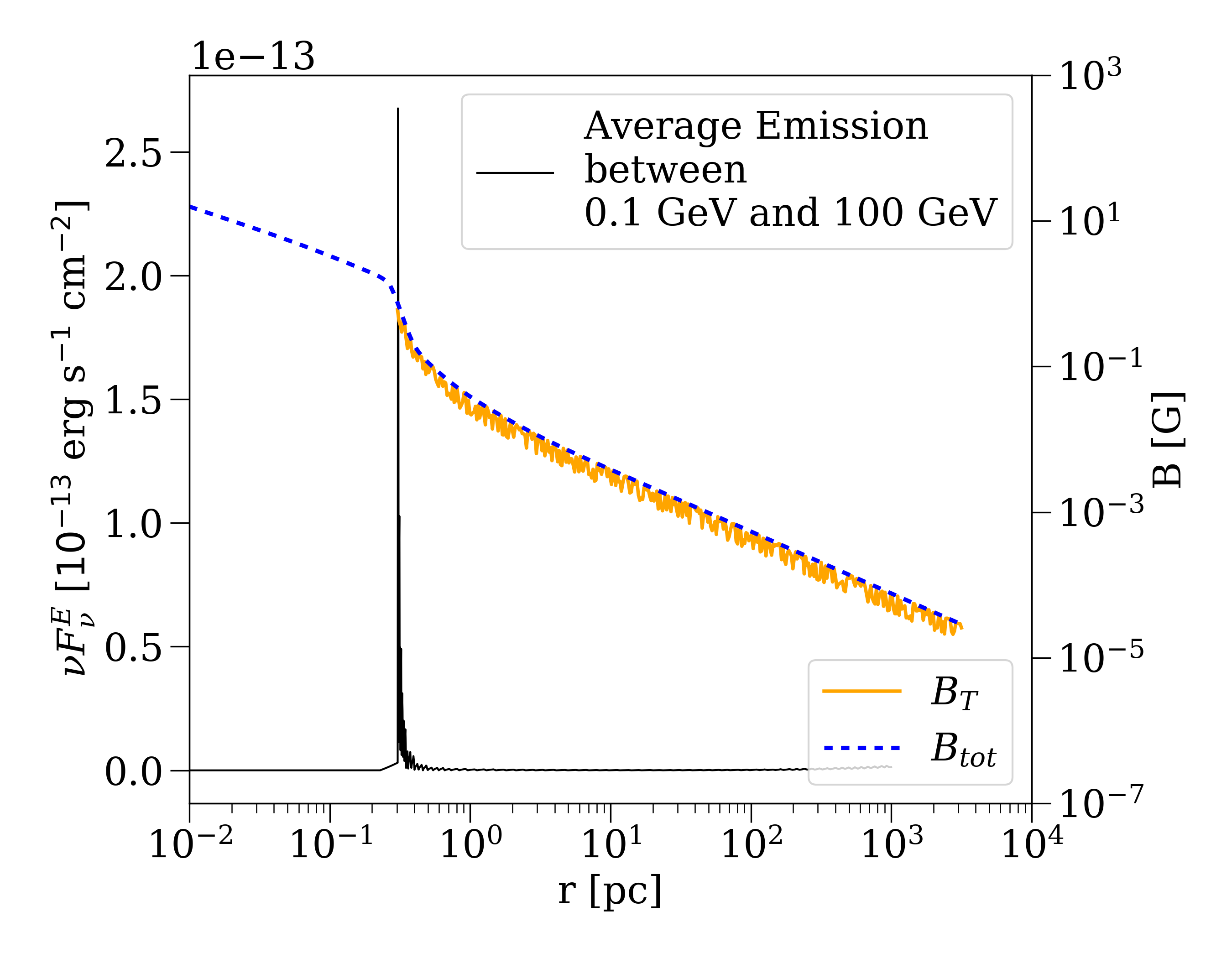}
  \caption{One realisation of the jet magnetic field ($\alpha=1$, $f=0.7$), showing the overall field strength (blue dashed) and the transverse magnetic field strength (orange, calculated from $r_{em}$ onward) used for calculating ALP-photon conversion, against $r$, the distance down the jet. $\nu F_{\nu}$ from the PC model is also shown, showing that a very localised gamma-ray emission region is reasonable.}
  \label{fig:b_vfv}
\end{figure}
\subsection{The importance of jet magnetic field structure}
\label{sec:structure}
Above, we have shown that the blazar jet is an important mixing region for possible ALP exclusions, and so, will have to be included for future searches. The analysis of Sec. \ref{sec:matter} was done using the simple jet model of Refs \cite{ronc_agn} and \cite{galanti_blazars}. As was shown in Sec. \ref{subsec:jetfield}, actual jet magnetic field structure is certainly more complicated and is not very well understood. It, therefore, makes sense to explore what effects a more complicated jet magnetic field structure can have on the mixing. That is to say that given that we need to include a jet model, we want to find out how detailed our model has to be. To this end, we parameterize a more detailed jet model (Sec. \ref{subsec:model}) and then investigate how the fits vary across this new magnetic field parameter space (Sec. \ref{subsec:varying}).
\begin{figure}
  \centering
    \includegraphics[width=0.5\textwidth]{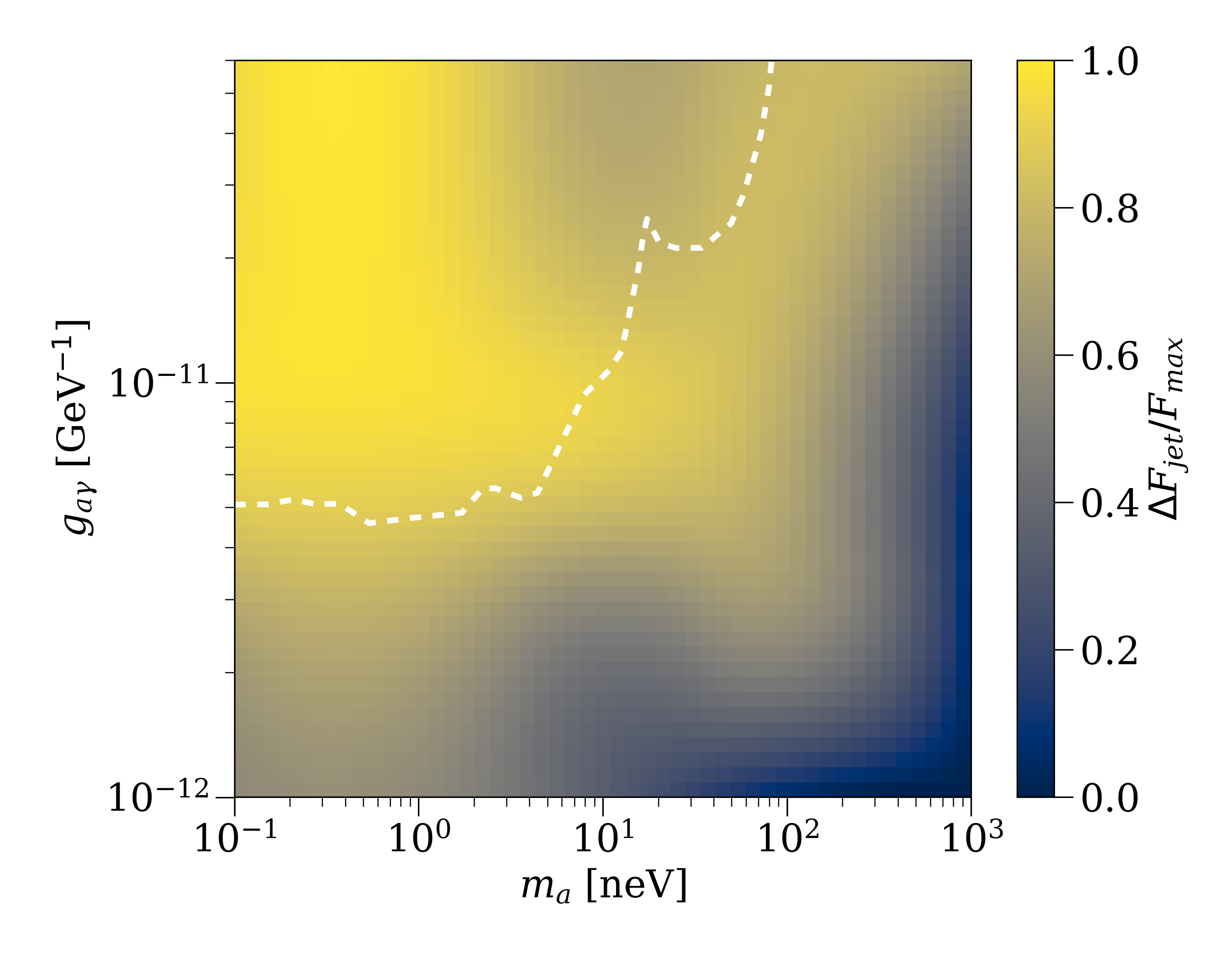}
  \caption{$\Delta F_{jet}/F_{max}$ over ALP mass-coupling parameter space. $\Delta F_{jet}$ is the change in $F_{jet}$ when the jet magnetic field parameters are varied within the limits shown in Table \ref{tab:params}. $F_{max}$ is the maximum value of $F_{jet}$ obtained during this variation. Dashed line shows current exclusions (see Sec. \ref{sec:preintro}).}
  \label{fig:deltas}
\end{figure}
\subsubsection{Jet model}
\label{subsec:model}
We compute our jet magnetic field in the jet rest frame. Practically, this means the jet-frame energy of an ALP-photon beam must be used while it is propagating through the jet, instead of its observed energy. This transformation is done by $E_{jet}=E_{obs}/\delta(r)$, where $\delta$ is the Doppler factor. For simplicity, we assume the viewing angle to be small ($\sim 1/\Gamma$), which gives $\delta(r)\approx\Gamma(r)$ where $\Gamma$ is the jet bulk Lorentz factor, which depends on $r$. Bulk Lorentz factors can be read from the PC model; the best fit PC model for Mrk 501 has $\Gamma(r_{em}) \sim 9$ and the deceleration from then on is logarithmic, $\Gamma(r>r_{em})\propto \log(r_{em}/r)$.
The overall magnetic field strength is taken from the PC model, consistent with observations: $\propto1/r^a$ in the parabolic base, and $\propto1/r$ in the conical jet with a smooth transition between the two. Figure \ref{fig:b_vfv} shows the overall B-field strength used. It also shows that the emission between 0.1 GeV and 100 GeV from the PC model is very localised at the transition region, which justifies the use of a single point as the gamma-ray emission region.
\par
Our jet magnetic field model is made up of a tangled component and a helical component that turns from poloidal (pointing down the jet) to toroidal (transverse) as it moves down the jet\footnote{\texttt{PYTHON} code of this model will be made publicly available by its inclusion in the \texttt{gammaALPs} package, which can be found at: \url{https://github.com/me-manu/gammaALPs}}.  This model fits with theory, observation, and simulations (Sec. \ref{subsec:jetfield}) -- and dealing with the tangled and helical components at this level of detail is not a feature of current models. We describe the helical component with two parameters, and the tangled component with one (see Table \ref{tab:params}). The rate at which the transverse component of the helical field changes with distance down the jet is governed by $\alpha$ ($B_T \propto r^{-\alpha}$) and the distance along the jet at which the helical component becomes toroidal is given by $r_T$. This is, of course, a simplification but allowing $\alpha$ to be different from $1$ can take into account the parabolic base and the fact that we do not know how fast the helical field turns from poloidal to toroidal.
The tangled component is a constant fraction of the total magnetic energy density, as in Ref. \cite{gabuzda_mrk501}:
\begin{equation}
    \frac{B^2_{tangled}}{B^2_{helical}}=\frac{f}{1-f}.
\end{equation}
Figure \ref{fig:b_vfv} also shows one realisation of the transverse magnetic field ($B_T$) for one set of field parameters ($\alpha=1$, $f=0.7$).\par
We now vary the parameters within realistic ranges to see how the jet structure affects the fits. We allow $\alpha$ to vary between $0.2$ and $1.5$, consistent with simulations (e.g. \cite{mckinney_2006}). As this transition from poloidal to toroidal is only expected down near the base, $r_T$ is varied between $0.1$ pc and $10$ pc. A constant pitch angle helical field could be consistent with observations out to kpc scales for powerful jets but for ALPs, which only see the transverse component, this is just equivalent to a weaker field. 
Following Ref. \cite{gabuzda_mrk501}, which fits $f$ to RM maps of Mrk 501, we take $f$ to vary between $0$ and $0.7$. The large range of possible values these parameters can take indicates how little we know about jet magnetic fields and hence, why their effect on ALP-photon mixing should be understood before a simplistic jet model is used to find ALP exclusions.

\begin{table}
 \caption{Parameters used for jet magnetic field structure model.}
  \centering
  \begin{tabular}{m{0.06\textwidth} m{0.3\textwidth} m{0.1\textwidth}}
    \toprule\toprule
    Name     & Description     & Values used \\
    \midrule\midrule
    $\alpha$ & Helical $B_T \propto r^{-\alpha}$  & 0.2 -- 1.5     \\
    \midrule
    $r_T$     & Radius at which helical field becomes toroidal & 0.1 -- 10 pc   \\
    \midrule
    $f$    & Fraction of magnetic energy density in tangled field       & 0 -- 0.7  \\
    \bottomrule\bottomrule
  \end{tabular}
  \label{tab:params}
\end{table}

\begin{figure*}
  \centering
    \includegraphics[width=1.0\textwidth]{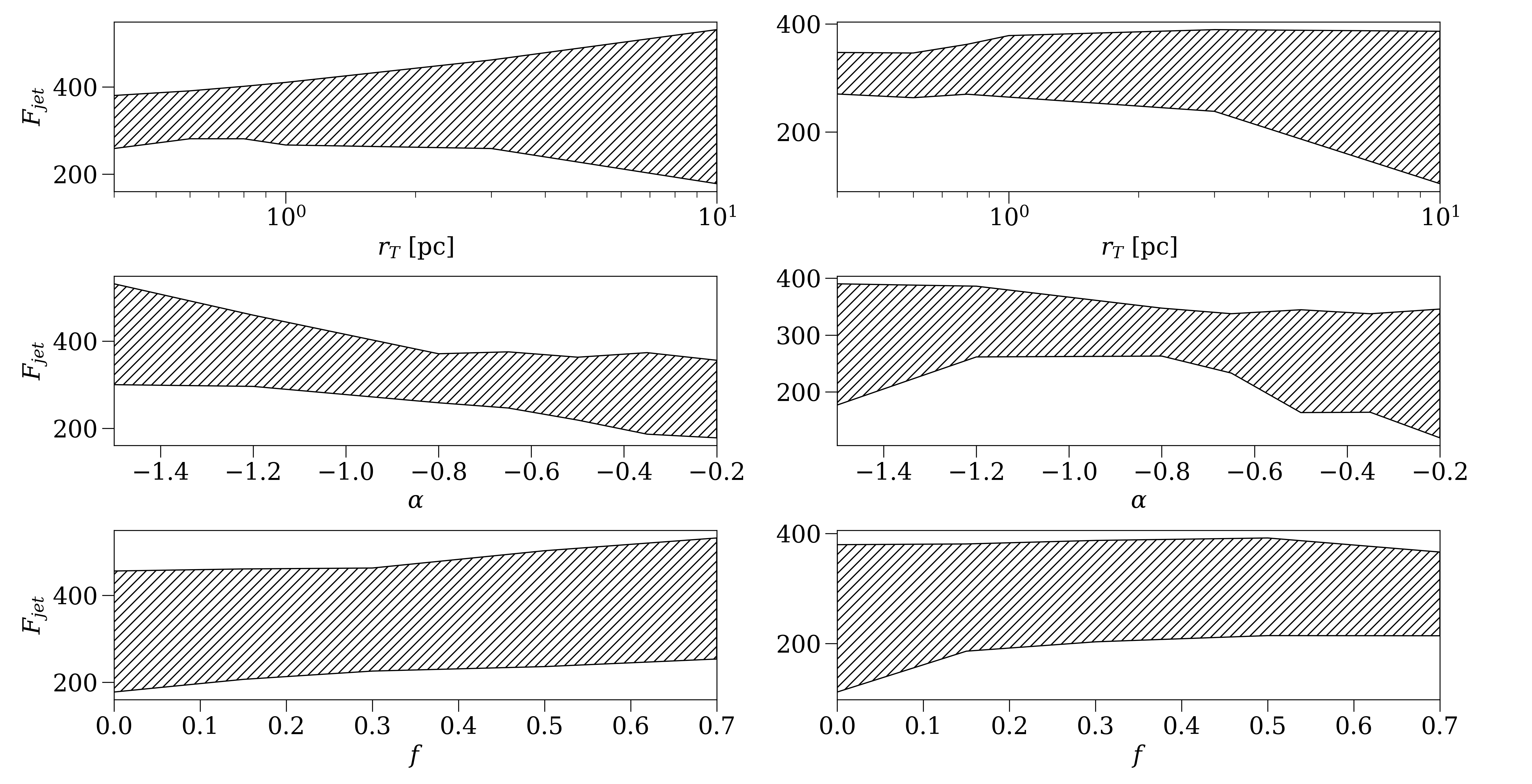}
  \caption{$F_{jet}$ vs magnetic field parameters for $m_a=1$ neV, $g_{a\gamma}=2\times10^{-12} \text{GeV}^{-1}$ (left column) and $m_a=100$ neV and $g_{a\gamma}=7\times10^{-12} \text{GeV}^{-1}$ (right column). Shaded regions show values of $F_{jet}$ that can be obtained by varying the other magnetic field parameters.}
  \label{fig:Fs_pars_both}
\end{figure*}

\subsubsection{Varying jet structure}
\label{subsec:varying}

Figure \ref{fig:deltas} shows $\Delta F_{jet}/F_{max}$ over the ALP $(m_a,g_{a\gamma})$ space, where $\Delta F_{jet}$ is the change in $F_{jet}$ (maximum minus minimum) running over the values of $r_T$, $\alpha$, and $f$ shown in Table \ref{tab:params}, and $F_{max}$ is the maximum value of $F_{jet}$ at that point. This shows how much reasonable changes in magnetic field structure can change the fits. As can be seen from Fig. \ref{fig:deltas}, varying the magnetic field parameters has a large effect on the fits: $\Delta F_{jet}/F_{max} > 0.3$ for almost all the ALP parameter space.

Fig. \ref{fig:Fs_pars_both} shows how each of the magnetic field parameters affect the fits for two choices of ALP parameters in the interesting unprobed region. The plots show $F_{jet}$ vs a given parameter ($r_T$, $\alpha$, $f$). The shaded bands show the range of values of $F_{jet}$ that can be produced by varying the other parameters while keeping the parameter in question fixed. For example, the band in $f$ is calculated by fixing $f$ at multiple points (e.g. 0., 0.2, 0.4, etc.), each time finding the maximum and minimum $F_{jet}$ that can be produced by changing the other parameters ($\alpha$ and $r_T$ in this case). These maxima and minima give the top and bottom of the band respectively. The bands therefore show how much a given parameter is contributing to the fits when it takes on any given value.  A broad shaded region means that that parameter is not strongly affecting the fits at that point -- i.e. changing the other parameters can change the fits a lot. A narrow shaded region means the opposite, that changing the other parameters does not alter the fits much because this parameter is dominating them. For both values of $(m_a,g_{a\gamma})$ in Fig. \ref{fig:Fs_pars_both}, the values of $r_T$ and $\alpha$ have a larger effect on $F_{jet}$ (narrower shaded regions) than $f$ does; and $r_T$ starts to have less of an effect as it increases above a few parsecs. Also, the fact that the bands in $r_T$ remain relatively horizontal at low values means that if a strong upper limit on $r_T\lesssim$ pc could be found, the overall values of $\Delta F_{jet}/F_{max}$ could be significantly reduced. That is to say that the structure of the helical component of the jet magnetic field right down at the base of the jet is making the strongest difference to the fits. This is a similar effect to what we saw earlier when comparing mixing in the jet with the IGMF (discussed in Sec. \ref{sec:matter}). The tangled jet component, $f$ has a similar effect on the spectra as the random domains of the IGMF, causing small oscillations of the spectrum around its large scale shape. The large scale shape is caused by the strong mixing that happens in the strongest magnetic fields at the jet base where the helical component matters.
\begin{figure*}
  \centering
    \includegraphics[width=\textwidth]{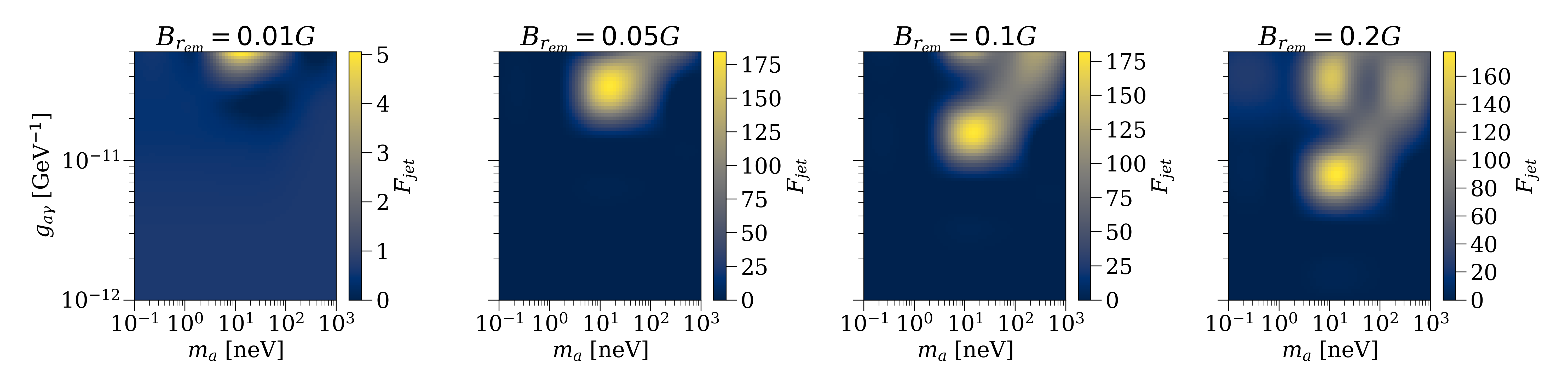}
  \caption{$F_{jet}$ for four different values of $B_{r_{em}}$. Below $B_{r_{em}} \sim 0.05$G the jet becomes unimportant.}
  \label{fig:4Bs}
\end{figure*}
In Sec. \ref{sec:matter} (Figs \ref{fig:fs} and \ref{fig:diffs}) we showed that for a specific set of reasonable jet parameters, the jet could be used as a mixing region to probe new ALP parameter space. By letting the magnetic field parameters vary over a reasonable range, in Fig. \ref{fig:deltas} we show that, in general, the jet could still be used to probe this region -- $F$ will tend to stay greater than zero. This is promising for future searches. However, the large values of $\Delta F_{jet}/F_{max}$ in Fig \ref{fig:deltas} mean that for any individual blazar, we would need to know the specific jet magnetic field parameters to accurately model the ALP spectrum and obtain these constraints; and at the moment, we do not know the specific jet magnetic field structure of individual sources in sufficient detail from VLBI measurements. One way of approaching this problem would be, at the cost of exclusion capability, to leave the jet magnetic field free in the fit. 

\subsection{Beyond Mrk 501}
\label{subsec:beyond}
We have only simulated spectra of Mrk 501. In order to extend our conclusions to other sources we must consider the intrinsic differences in jets and the differences in their environments. Firstly, the values of magnetic field parameters used (Table \ref{tab:params}) and the reasons for using them are not only applicable to Mrk 501 but to any similar BL Lac type object. It is not only the structure of the magnetic field in Mrk 501 that we do not know precisely, but that of jets in general. The fact that the same PC model including a highly magnetized base fits so many blazar spectra suggests that the basic structure and mechanisms within other blazars should be similar. This means that, in general, if mixing within the jet is important for a source, the magnetic field structure of the source will need to be known more precisely than current models or observations can determine. Extending beyond Mrk 501 is, therefore, a question of whether intrinsic or environmental differences in other sources prevent the jet from being an important mixing region. 
The relevant intrinsic properties that vary between sources are the field strength, the location of the emitting region, the jet length, and the viewing angle. Changing the viewing angle is essentially just changing the length of the jet that the ALP sees before passing out of it (it will also change the Doppler factor, which will shift the ALP spectrum in energy). The $1/r$ dependence of the overall field strength down the length of the jet described in Sec. \ref{subsec:jetfield} applies to jets in general. As shown in Sec. \ref{sec:matter} it is mainly the mixing in the strong magnetic field at the base of the jet that makes the jet an important mixing region. Indeed, excluding regions of the jet after the field has dropped below $~10^{-4}$G doesn't change the $P_{\gamma\gamma}$'s very much at all. Therefore, changing the length of the jet does not matter much as it amounts to lengthening or shortening the region of weak field at the end of the jet, which is unimportant. The PC model provides emission information all down the jet but finds in all cases that gamma-ray emission is strongly dominated by the transition region between the parabolic base and conical ballistic jet. Changing the location of the gamma-ray emission region is, therefore, equivalent to changing the magnetic field at the emission region. Therefore the only intrinsic property that matters is the magnetic field strength at the transition (emission) region. For Mrk 501, the PC best-fit value is $0.8$G. This is the value used in Sec. \ref{sec:matter} for finding the importance of the jet as a mixing region. In Ref. \cite{pc_nc} Potter $\&$ Cotter fit the blazar spectra of 42 blazars with the PC model. The largest best-fit value they found was $2.91$ G (J2143.2+174 / S3 2141+17), and the smallest was $0.0057$ G (J1504.3+1030 / PKS1502+106).

Figure \ref{fig:4Bs} shows the values of $F_{jet}$ for 4 values of the magnetic field at the emission region ($B_{r_{em}}$) between $0.01$ G and $0.2$ G. Decreasing $B_{r_{em}}$ increases the lowest value of $g_{a\gamma}$ for which mixing in the jet is important. This makes sense: With a stronger field you can probe weaker couplings. As $B_{r_{em}}$ drops below $0.05$ G, however, $F_{jet}$ quickly becomes very small everywhere; mixing ceases to occur at all for ALPs in this parameter space (note the different colourbar for $B_{r_{em}}=0.01$ G). This means that jets with magnetic fields at the emission region $B_{r_{em}}< 0.05$ G are not important mixing regions for ALPs. These sources could then be used for ALP searches without requiring a detailed knowledge of the jet magnetic field structure. $0.05$ G is a very weak magnetic field strength for the emission region of a BL Lac (see, e.g. \cite{osullivan_gabuzda}), but could be reasonable for many FSRQs (and is above the best fit for many of the blazars fitted in PC \cite{pc_nc} or from estimates from radio core-shift measurements, e.g. \cite{zam}). \par
\begin{figure}
  \centering
    \includegraphics[width=0.48\textwidth]{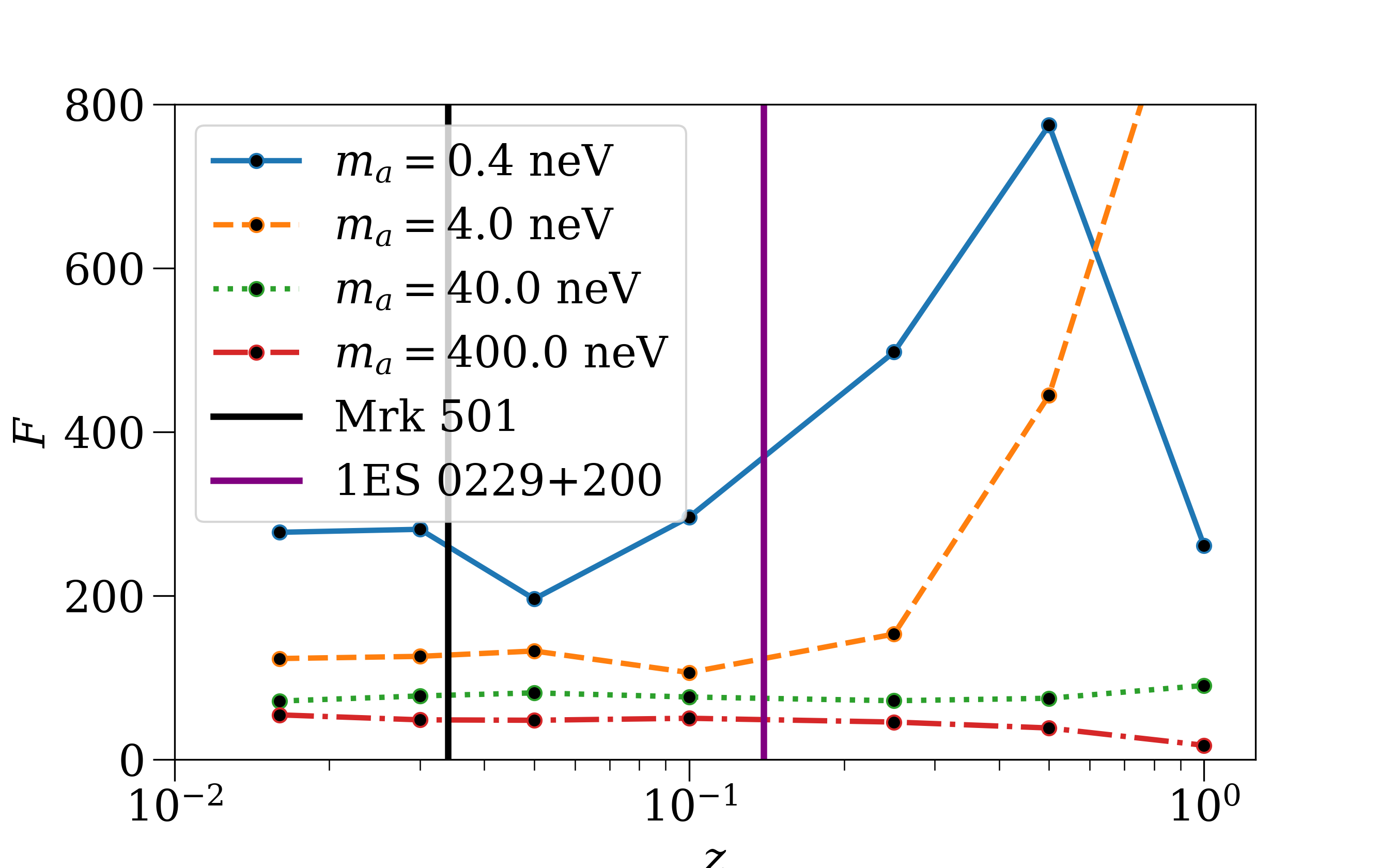}
  \caption{Redshift dependence of $F$ for different values of $m_a$ at coupling strength $g_{a\gamma}=1.5\times10^{-11}\text{GeV}^{-1}$. Fig. \ref{fig:diffs} shows that the jet starts to dominate the fits at around $m_a \gtrsim 10$ neV, which is where $F$ ceases to change with $z$. Redshifts of blazars Mrk 501 (black) and 1ES 0229+200 (purple) are shown for reference; these are the two blazars discussed in the context of CTA ALP searches in Ref. \cite{galanti_blazars}.}
  \label{fig:zdep}
\end{figure}

Of course, if these searches would like to probe new parameter space then a mixing region other than the jet must be used. Fig. \ref{fig:zdep} confirms the fairly obvious result that changing the redshift of the source only affects the fits for regions of ALP parameter space where the IGMF is already dominant; the regions of $(m_a,g_{a\gamma})$ space where the jet dominates remain unaffected by changes in redshift. This means that even for sources with larger $z$ (e.g. 1ES 0229+200 considered in Ref. \cite{galanti_blazars} for CTA searches) the IGMF cannot be used to probe new parameter space.

\begin{figure}
  \centering
    \includegraphics[width=0.5\textwidth]{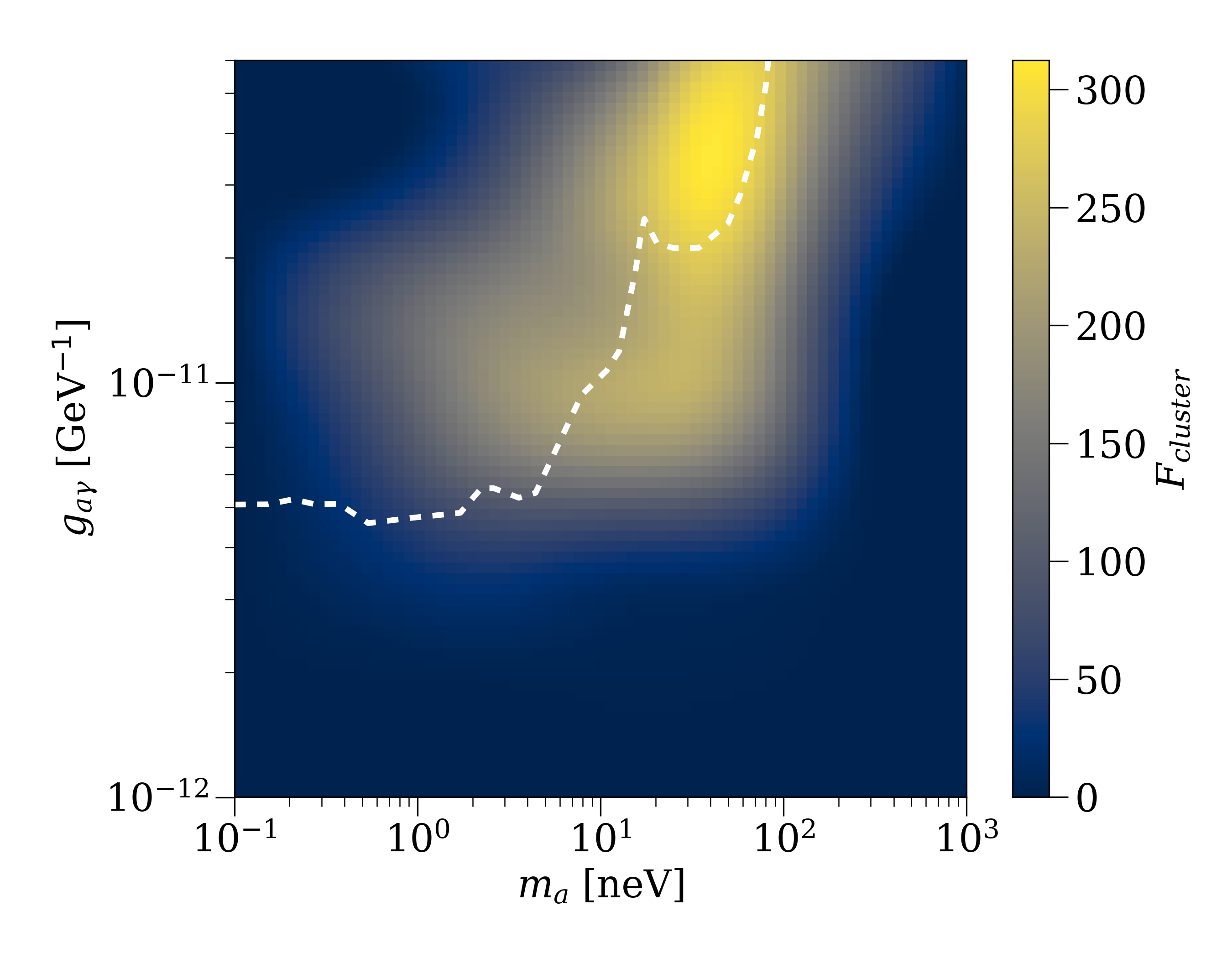}
  \caption{$F_{cluster}$ for cool core cluster parameters as used in Ref. \cite{manuelfermi}, with $B_{rms}=10 \mu$G. Dashed line shows current exclusions (see Sec. \ref{sec:preintro}).}
  \label{fig:cluster}
\end{figure}

One option is to use a source that is located in a highly magnetized cluster. This has been done in Refs. \cite{manuelfermi} and \cite{hess_exclusions} to obtain ALP exclusions with the Fermi-LAT and H.E.S.S. spectra of NGC1275 (in the Perseus cluster) and PKS2155-304 respectively. Faraday rotation measurements of cool core galaxy clusters like Perseus suggest magnetic fields that are turbulent, with strengths of around $10 \mu$G. Figure \ref{fig:cluster} shows $F_{cluster}$ for one realisation of the cluster field model used in Ref. \cite{manuelfermi}, with the parameters shown in their Table 1. As can be seen from the figure, the cluster can provide an important mixing region for a large region of ALP parameter space, much of which is beyond current constraints. This is because the cluster has a roughly $10 \mu$G magnetic field (much stronger than the IGMF), over a large distance ($500$ kpc which is longer than the jet), so that it can produce a mixture of small- and large-scale oscillations in the energy spectra. \par
\begin{figure}
  \centering
    \includegraphics[width=0.5\textwidth]{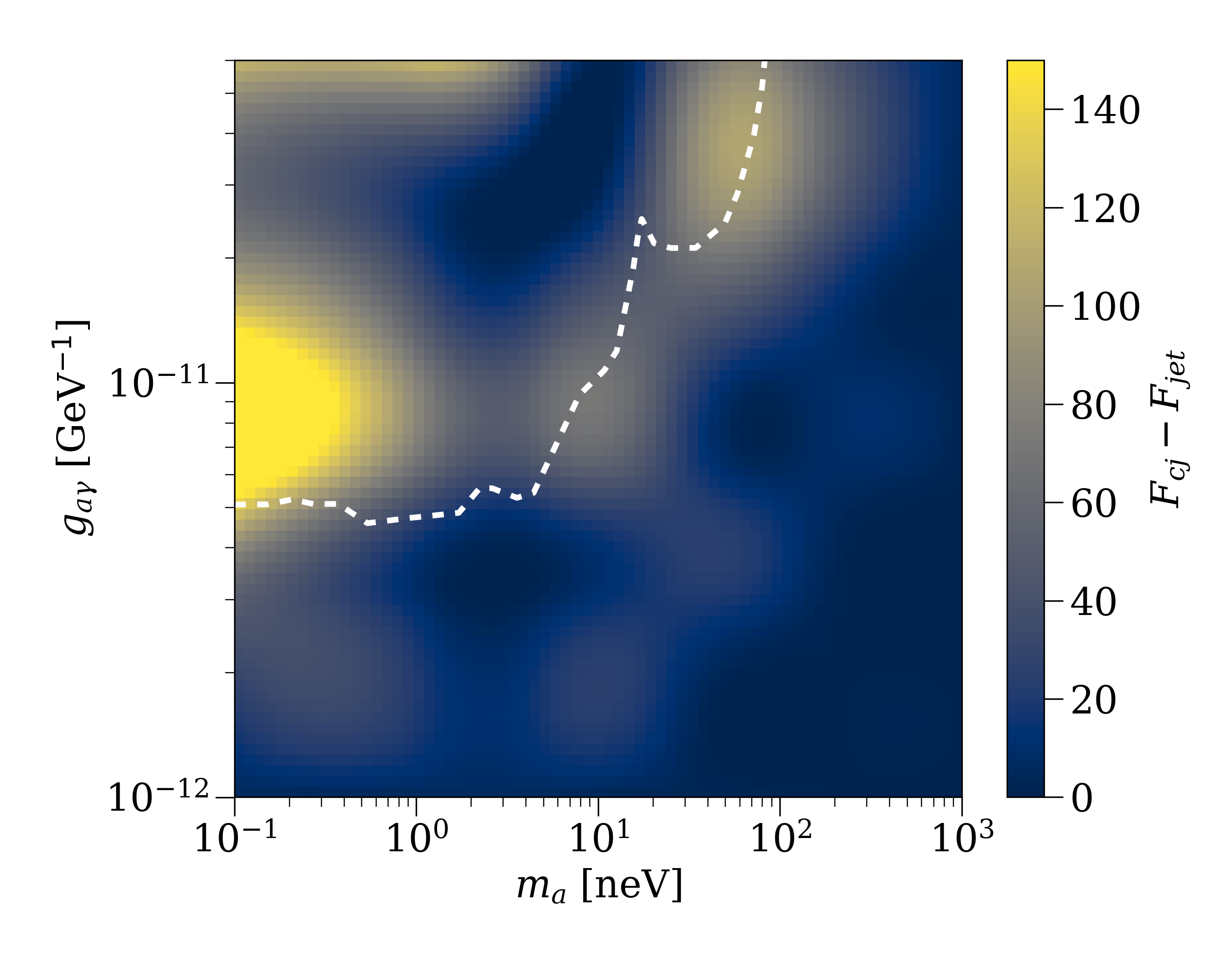}
  \caption{$F_{cj} - F_{jet}$ for cool core cluster parameters as used in Ref. \cite{manuelfermi}, with $B_{rms}=10 \mu$G. $F_{cj}$ is the value of $F$ when the beam is propagated through the jet and the cluster, and $F_{jet}$ when it is propagated through just the jet. Dashed line shows current exclusions (see Sec. \ref{sec:preintro}).}
  \label{fig:cj-j}
\end{figure}

In fact, as shown in Figure \ref{fig:cj-j}, mixing in the cluster dominates even when the jet is highly magnetized. Figure \ref{fig:cj-j} shows $F_{cj} - F_{jet}$, where $F_{cj}$ is the value of $F$ when the ALP-photon beam is propagated through both the jet and one cluster realisation. Unlike in Figure \ref{fig:diffs} ($F-F_{jet}$), $F_{cj} - F_{jet}$ does not become very small over the region of unprobed parameter space. Mixing in the jet will change the initial conditions for mixing in the cluster, and so will slightly change the exact form of an ALP spectrum for an individual realisation of a cluster magnetic field (hence $F_{cj} - F_{jet}$ doesn't look exactly like $F_{cluster}-F_{jet}$) but not enough to be disentangled from simply another realisation of pure cluster mixing. This means that the jet does not contribute strongly for searches using an ensemble of cluster realisations as has been used to get the current exclusions.  Therefore, sources located near the center of magnetized clusters (like NGC1275 in the Perseus Cluster used in Ref. \cite{manuelfermi}) can be used to search for ALPs without requiring a detailed model of the jet. In particular, the current important constraints from Fermi-LAT and H.E.S.S. are still valid, and the same method could be used to probe farther ALP parameters in the future (e.g. with CTA as discussed in Ref. \cite{cta_gpropa}). These searches require a model of the cluster field, changes in which can also affect the mixing (see Ref. \cite{troitsky}), and which gets harder to put observational constraints on as the redshift increases. Indeed, there are many sources for which the cluster environment is unknown. However, blazars observed more accurately at a higher redshift in the future, by CTA for example, are likely to be of the brighter quasar type. These have been shown by radio and x-ray studies to be more likely to reside in less rich and, therefore, less magnetized cluster environments \cite{croston,gendre}. This is probably because of two factors. Firstly, for a given jet power a denser environment is more likely to disturb the jet. Secondly, environmental differences can affect the accretion mode of the central engine, with low excitation objects in denser environments being fed by hot ICM gas and high excitation objects in less dense environments being fed by a traditional cooler accretion disk \cite{Heckman_and_Best}. This could mean that in the future, our ignorance of cluster field strengths and configurations could be less important, as mixing in weaker clusters is less dominant and that mixing in the jet will, therefore, be crucial to these future searches. The dominance of the cluster field would also be less for sources closer to the edge of their cluster, even if it is highly magnetized.

Another option is to use the lobes of powerful FSRQ sources as discussed in Ref. \cite{ronc_agn}. They show that, for ALP masses $m_a < $ neV, efficient mixing in the lobes can wipe out oscillations from the jet and cause equipartition between the ALP and photon states -- a constant conversion probability independent of energy. While the lack of spectral oscillations is not good for ALP searches, it does mean that our lack of knowledge about the jet magnetic field structure is not important for these sources when searching in this parameter range.

\subsection{ALPs as probes}
\label{subsec:probes}
It is worth pointing out that next generation experimental axion searches (particularly IAXO \cite{iaxo}) will be covering the same region of ALP parameter space for which mixing in the jet is important (cf. Figs \ref{exclusionplot} and \ref{fig:fs}). This means that, in the event of a discovery, ALPs could be used as probes of jet magnetic field structure. If the values of $m_a$ and $g_{a\gamma}$ are known, a sort of reversed ALP search could be performed where the magnetic field parameter space is scanned as opposed to the ALP space, allowing certain magnetic field parameters to be excluded as opposed to ALP parameters. This kind of process could potentially be used to identify the emission region of gamma-ray flares in blazars, the subject of ongoing debate (e.g., \cite{manuel_flares}). One difficulty of this process (other than the obvious difficulty of ALPs having to be discovered first) involves the oscillatory nature of ALP-photon mixing. For example, there is a degeneracy between possible emission regions, as shifts of one oscillation length reproduce almost exactly the same spectrum.
This could also be done to investigate cluster magnetic fields, particularly around distant blazars whose cluster environment is difficult to constrain with other observations.

\section{Conclusions}
\label{sec:conclusions}
We have investigated the relevance of jet magnetic field structure for blazar ALP searches. ALP spectra of Mrk 501 have been simulated by propagating ALP-photon beams through all the magnetic field environments along the line of sight. By comparing fits to ALP-less spectra with and without mixing in the jet for a specific set of reasonable jet field parameters, we have shown that the jet is an important region for ALP-photon mixing. This is because mixing in the strong magnetic field of the jet tends to put the mixing equations into the strong-mixing regime for a large range of energies, which sets the large scale structure of the spectrum. Mixing in the IGMF, however, tends to produce small scale oscillations around this shape.\par
In particular, mixing in jets is important for the unprobed ALP parameter space likely to be the target exclusion region for future blazar ALP searches. Or, in other words, mixing in the jet could enable us to probe new regions of the ALP parameter space. It is, therefore, important to investigate how variations in the jet magnetic field configuration affect the ALP-photon mixing in jets, to see how detailed our jet magnetic field models have to be for these searches, and whether current observational and theoretical constraints enable us to build such a model. We have done this by imposing a reasonable jet field structure onto the PC jet framework. Our model is composed of a helical component that transitions from poloidal to toroidal, and a tangled component. The field structure can be described with three parameters: $f$, the fraction of magnetic field energy density in the tangled component; $\alpha$, the power law index of the transverse component of the helical field; and $r_T$, the distance along the jet at which the helical component becomes toroidal. By scanning feasible magnetic field parameter space (i.e. allowing these three parameters to vary within observational and theoretical limits), we have shown that reasonable changes in jet magnetic field structure, particularly the helical component at the highly magnetized base of the jet, have a large effect on ALP-photon mixing and hence on possible exclusions. Changing the magnetic field parameters rarely reduces the mixing to zero, so the jet will still be able to probe new ALP parameter space. The strong dependence of the fits on the specific magnetic field parameters does mean, however, that future ALP searches that want to probe the unexcluded region of ALP parameter space, where mixing in the jet is important, require a detailed model of the specific jets they are looking at to reliably exclude ALP parameters with blazar spectra. This will require better knowledge of jet magnetic fields than we currently have, or enough computing power to marginalize over magnetic field parameter space. We have shown that these results are applicable beyond Mrk 501, to any blazar with a magnetic field at its emission region larger than about $0.05 G$. This is a very weak field for BL Lac type objects, but could be reasonable for some FSRQs \cite{zam}. In the case of a weaker field, our ignorance of jet structure is unimportant, but another mixing region between us and the source would then be required to probe new ALP parameter space. We have demonstrated that, for sources located in a highly magnetized cluster, the intracluster magnetic field is capable of wiping out the effects of jet mixing. This means that the current limits set on ALP parameter space from gamma-ray spectra are still valid, because they have used the Perseus Cluster as their mixing region. Our results are particularly important for the upcoming ALP searches planned with CTA. It has also been pointed out that this could work the other way around, with ALPs acting as probes of jet structure in the event of their discovery.

\section*{Acknowledgements} \label{sec:acknowledgements}
J. D. acknowledges an STFC Ph.D. studentship.  M. M. acknowledges the research leading to these results has received funding from  the  European  Union's  Horizon 2020 research and innovation program under the Marie Sk{\l}odowska-Curie Grant agreement GammaRayCascades No~843800. G. C. acknowledges support from STFC grants ST/S002618/1 and ST/M00757X/1, and from Exeter College, Oxford.

\appendix
\section{ALP-Photon Mixing}
\label{app:mixing}
Including the ALP-photon mixing term, ALPs can be described by the Lagrangian
\begin{equation}\label{eq:fulllagrangian}
    \Lik_{ALP} = \frac{1}{2}\partial^{\mu}a\partial_{\mu}a - \frac{1}{2}m_{a}^{2}a^{2} + \Lik_{a\gamma}.
\end{equation}
Mixing between ALPs and photons in an external magnetic field is described by the Lagrangian $\Lik_{a\gamma}$, given in Equation \ref{eq:mixlagrangian}. As the electric field ($\boldsymbol{E}$) of a photon is orthogonal to its wave vector ($\boldsymbol{k}$), the dot product in $\Lik_{a\gamma}$ means only the component of the external magnetic field ($\boldsymbol{B}$ in $\Lik_{a\gamma}$) that is transverse with respect to the direction of beam propagation ($\boldsymbol{B}_{T}$) is important for mixing. Similarly, only the component of $\boldsymbol{E}$ in the plane spanned by $\boldsymbol{B}$ and $\boldsymbol{k}$ is involved in the mixing ($\boldsymbol{E}_{||}$), meaning only photon polarisation states along $\boldsymbol{E}_{||}$ mix with ALPs \cite{ronc_big}. If we consider a linearly polarised photon beam of energy $E$ (from now on $E$ is used for energy and not electric field) propagating in the $y$ direction in a homogeneous magnetic field, it follows from Eq. \ref{eq:fulllagrangian} that the propagation of the ALP-photon beam evolves according to the second-order wave equation:
\begin{equation}\label{eq:2ndorderwe}
    \Big(\frac{d^{2}}{dy^{2}} + E^{2} + 2E\mathcal{M}_{0}\Big)\Psi(y) = 0.
\end{equation}
Here $\Psi(y)\equiv (A_{x}(y),A_{z}(y),a(y))^{T}$ where $A_{x}$ and $A_{z}$ are the photon amplitudes with polarisation in the (arbitrarily chosen, but orthogonal to $y$ and each other) $x$ and $z$ directions and $a(y)$ is the amplitude associated with the ALP. $\mathcal{M}_{0}$ is the ALP-photon mixing matrix. As we are considering a homogeneous magnetic field and propagation in regions where the refractive index satisfies $|n-1|\ll1$, the following relations hold \cite{raffstod}:
\begin{dmath}
    \Big(\frac{d^{2}}{dy^{2}} + E^{2}\Big)\Psi(y) = \Big(i\frac{d}{dy} + E\Big)\Big(-i\frac{d}{dy} + E\Big)\Psi(y) = 2E\Big(i\frac{d}{dy} + E\Big)\Psi(y).
\end{dmath}
Equation \ref{eq:2ndorderwe} can, therefore, be linearised into the Schr{\"o}dinger-like equation,
\begin{equation}\label{eq:1storderwe}
    \Big(i\frac{d}{dy} + E + \mathcal{M}_{0}\Big)\Psi(y) = 0.
\end{equation}
Choosing the $z$ direction to be aligned with $\boldsymbol{B}_{T}$ (from now on, $\boldsymbol{B}_{T}=B\boldsymbol{\hat{z}}$) and neglecting Faraday rotation (because of the high gamma-ray energies we are considering) the mixing matrix $\mathcal{M}_{0}$ takes the form,
\begin{equation}\label{eq:mm1}
    \mathcal{M}_{0} = 
    \begin{pmatrix}
    \Delta_{\gamma\gamma} & 0 & 0 \\
    0 & \Delta_{\gamma\gamma} & \Delta_{a\gamma}\\
    0 & \Delta_{a\gamma} & \Delta_{aa}
    \end{pmatrix}.
\end{equation}
Where the diagonal terms arise from the propagation of photons and ALPs and the off diagonal terms are due to ALP-photon mixing. The photon term, $\Delta_{\gamma\gamma}$ includes a plasma contribution as well as contributions due to the QED vacuum polarisation effect, scattering off the CMB, and any absorption \cite{ronc_smoothed}:
\begin{equation}\label{eq:dgg}
    \Delta_{\gamma\gamma} = \Delta_{pl} + \Delta_{QED} + \Delta_{CMB} + \frac{i}{2\lambda_\gamma}.
\end{equation}
The plasma contribution depends on the effective mass of the photon in the plasma, $m_T$ (which is just the plasma frequency in a cold plasma: $\omega_{pl} \sim 0.037\sqrt{n_{e}}$ neV with the electron density $n_{e}$ in units of $\text{cm}^{-3}$),
\begin{equation}\label{eq:dpl}
    \Delta_{pl} = -\frac{m_{T}^{2}}{2E}.
\end{equation}
Inside a relativistic jet, we are not dealing with a cold thermal plasma, and so, the photon effective mass ($m_T$) will not simply be the plasma frequency (see \cite{raff_starlabs}). In this case, we can use the fact that the electron distribution function is a power law with an index related to the observed synchrotron power law index to find the effective mass of the photon in the plasma:
\begin{equation}\label{eq:mT}
    m_T^2 = \frac{\alpha}{\pi^2}\frac{(\beta - 1)}{m_e^{(\beta-1)}}n_{e}\int_{m_e}^{E_{max}} \frac{E^{-\beta}}{\sqrt{E^2 - m_e}} dE,
\end{equation}
where $\alpha$ is the fine structure constant, $\beta$ is the electron distribution function index ($f(E)dE \propto E^{-\beta}dE$ with $J(\nu)d\nu \propto \nu^{\frac{(1-\beta)}{2}}d\nu$), and $E_{max}$ is the maximum electron energy in the population. $m_T$ can be quite different to $w_{pl}$, depending on the electron energies and the index, $\beta$. However, for gamma-ray photon energies and the magnetic field strengths in jets the QED vacuum polarisation term is much more important than the plasma term regardless. For our jet, $n_e\propto r^{-2}$ with $n_{e}^{em}=5\times10^{4}\text{cm}^{-3}$ at the emission region as in Ref \cite{galanti_blazars}. Because $m_T$ is a function of $n_e$, it is not constant, but changes along the jet as a function of distance.\par
The QED vacuum polarisation term depends on the magnetic field as,
\begin{equation}\label{eq:dQED}
    \Delta_{QED} = \frac{\alpha E}{45\pi}\Big(\frac{B}{B_{cr}}\Big)^{2},
\end{equation}
with $\alpha$ the fine structure constant and $B_{cr}$ the critical magnetic field $B_{cr}=m_{e}^{2}/|e|\sim 4.4\times10^{13}G$. The CMB term is constant and has been calculated to be \cite{dobrynina_raffelt_cmb}:
\begin{equation}
    \Delta_{CMB} = 0.522\times 10^{-42} E \text{ eV}. 
\end{equation}
The term $i/2\lambda_{\gamma}$ takes into account any gamma-ray absorption processes with mean free path $\lambda_{\gamma}$.
The propagation term for the ALPs is,
\begin{equation}\label{eq:daa}
    \Delta_{aa}=-\frac{m_{a}^{2}}{2E},
\end{equation}
and the ALP-photon mixing term reads,
\begin{equation}\label{eq:dag}
    \Delta_{a\gamma}=-\frac{g_{a\gamma} B}{2}.
\end{equation}
Ignoring absorption, these equations of motion lead to ALP-photon oscillations with wave number (e.g. \cite{ronc_smoothed}),
\begin{dmath}\label{eq:dosc}
    \Delta_{osc}=\sqrt{\Delta_{\gamma\gamma}-\Delta_{aa})^{2} + 4\Delta_{a\gamma}^{2}} = \sqrt{
    \begin{aligned}
    \Big(\frac{|m_{a}^{2} - m_{T}^{2}|}{2E} + E\Big[\frac{\alpha}{45\pi}\Big(\frac{B}{B_{cr}}\Big)^{2} + \Delta_{CMB}\Big]\Big)^{2} \\+(g_{a\gamma}B)^{2}
    \end{aligned}
    },
\end{dmath}
which broadly explains why ALPs could lead to oscillatory features in gamma-ray spectra (flux vs energy) as the oscillation length between photons and ALPs, for a given environment ($B$ and $m_T$), depends on the beam energy. This means that the probability of the ALP-photon beam being in the ALP state once it has passed through the magnetic field environment will oscillate in energy. This oscillatory behaviour will be particularly prevalent when either of the first two terms ($\propto 1/E$ and $\propto E$) become similar to the mixing term, as then the oscillation length will depend strongly (and linearly) on energy. This gives two so-called critical energies, around which we expect strong oscillations in energy spectra:
\begin{equation}\label{ecrit}
    E_{cr}^{lower} = \frac{|m_{a}^{2}-m_{T}^{2}|}{2g_{a\gamma}B},
\end{equation}
and
\begin{equation}
    E_{cr}^{higher} = g_{a\gamma}B\Big[\frac{\alpha}{45\pi}\Big(\frac{B}{B_{cr}}\Big)^{2} + \Delta_{CMB}\Big]^{-1}.
\end{equation}
Similarly, when the third term in equation \ref{eq:dosc} dominates (the strong mixing regime) mixing becomes independent of energy and so will not produce energy-spectra oscillations.
While the treatment of a linearly polarised beam is useful for displaying this oscillatory behaviour, in gamma-ray astronomy we cannot measure photon polarisation, and so, we must treat the beam as unpolarised. Equation \ref{eq:1storderwe} must, therefore, be reformulated in terms of the density matrix $\rho(y) = \Psi(y)\Psi(y)^{\dagger}$, which gives the Von-Neumann-like equation:
\begin{equation}\label{eq:vonneu}
    i\frac{d\rho}{dy}=[\rho,\mathcal{M}_{0}].
\end{equation}
It is also more useful in practice to consider a magnetic field that makes an angle $\psi$ with $\boldsymbol{\hat{z}}$ as opposed to being aligned with it. This alters the mixing matrix by a similarity transform, giving
\begin{equation}\label{eq:mm2}
    \mathcal{M}_{0} = 
    \begin{pmatrix}
    \Delta_{\gamma\gamma} & 0 & \Delta_{a\gamma}\sin(\psi) \\
    0 & \Delta_{\gamma\gamma} & \Delta_{a\gamma}\cos(\psi)\\
    \Delta_{a\gamma}\sin(\psi) & \Delta_{a\gamma}\cos(\psi) & \Delta_{aa}
    \end{pmatrix}.
\end{equation}
Equation \ref{eq:vonneu} can be solved using the transfer matrix $\mathcal{U}(y,y_{0};E)$ associated with Eq. \ref{eq:1storderwe} ($\Psi(y) = \mathcal{U}(y,y_{0};E)\Psi(y_{0})$):
\begin{equation}\label{eq:solverho}
    \rho(y) = \mathcal{U}(y,y_{0};\psi;E)\rho(y_{0})\mathcal{U}^{\dagger}(y,y_{0};\psi;E),
\end{equation}
which gives the probability of a state $\rho_{1}$ being found in state $\rho_{2}$ after propagating from $y_{0}$ to $y$ as,
\begin{equation}\label{eq:1prob}
    P_{\rho_{1}\to\rho_{2}}(y)= Tr(\rho_{2}\mathcal{U}(y,y_{0};\psi;E)\rho_{1}\mathcal{U}^{\dagger}(y,y_{0};\psi;E)).
\end{equation}
We are not dealing with a single domain of constant field. For our case, the field environments have to be sliced into $N$ consecutive domains of constant $B$ and $\psi$, within which, the propagation equations can be solved exactly. For the jet we use 400 logarithmically spaced domains, which we found to be small enough slices not to affect the overall survival probabilities. The IGMF is already organised into domains of constant $B$ and $\psi$ in our model; for the GMF, we use 100 linearly spaced domains, and for the CMF, we use 4500 logarithmically spaced domains. From Eq. \ref{eq:1prob} the probability of a photon surviving, in any polarisation, the propagation through these $N$ consecutive domains is (as in Ref. \cite{manuel2}),
\begin{dmath}\label{eq:pgg}
    P_{\gamma\gamma} = Tr((\rho_{11}+\rho_{22}) \mathcal{U}(y_{N},y_{1};\psi_{N,...,1};E)\rho(0)\mathcal{U}^{\dagger}(y_{N},y_{1};\psi_{N,...,1};E)),
\end{dmath}
where $\rho_{11} =$ diag$(1,0,0)$ and $\rho_{22} =$ diag$(0,1,0)$ (the two photon polarisation states) and,
\begin{equation}\label{eq:proptans}
    \mathcal{U}(y_{N},y_{1};\psi_{N,...,1};E) = \prod_{i=1}^{N}\mathcal{U}(y_{i+1},y_{i};\psi_{i};E).
\end{equation}
If the initial beam is unpolarised, $\rho(0) = 1/2$diag$(1,1,0)$. The explicit expression of the transfer matrices used to find the photon survival probabilities ($P_{\gamma\gamma}$) can be found in Ref. \cite{ronc_big}.

\end{document}